\documentclass[12pt, draftclsnofoot, onecolumn]{IEEEtran}
\usepackage{cite}
\usepackage{url}
\usepackage{amsmath,amssymb,amsfonts}
\usepackage{algorithmic}
\usepackage{graphicx}
\usepackage{lipsum}
\usepackage{hhline}
\usepackage{tabularx}
\usepackage[table]{xcolor}
\usepackage{multirow}
\usepackage{subcaption}
\usepackage[font={small}]{caption}
\usepackage{dblfloatfix}    

\graphicspath{{graphics/}}
\usepackage{algorithm}
\usepackage{algorithmic}
\usepackage{verbatim}
\usepackage{cprotect}
\usepackage{afterpage}
\usepackage[export]{adjustbox}

\usepackage{siunitx}
\usepackage{booktabs}
\usepackage{makecell}
\usepackage{tabularx,booktabs}

\DeclareMathOperator*{\argmin}{arg\,min} 

\definecolor{LightCyan}{rgb}{0.88,1,1}

\def\BibTeX{{\rm B\kern-.05em{\sc i\kern-.025em b}\kern-.08em
		T\kern-.1667em\lower.7ex\hbox{E}\kern-.125emX}}
\usepackage{balance}

\begin{document}
	
	\title{Self-Supervised and Invariant Representations for Wireless Localization}
	
	\author{
		Artan Salihu$^\dagger$$^\ddagger$, Stefan Schwarz$^\dagger$$^\ddagger$ and Markus Rupp$^\dagger$
		\thanks{$^\dagger$ Institute of Telecommunications, Technische Universit{\"a}t (TU) Wien\\
			$^\ddagger$ Christian Doppler Laboratory for Dependable Wireless Connectivity for the Society in Motion\\
			Email: \{artan.salihu,stefan.schwarz,markus.rupp\}@tuwien.ac.at}}
	
	\markboth{Journal of \LaTeX\ . Submitted on January, 2023.}%
	{Shell \MakeLowercase{\textit{et al.}}: A Sample Article Using IEEEtran.cls for IEEE Journals}
	
	\maketitle
	
	\begin{abstract}
		In this work, we present a wireless localization method that operates on self-supervised and unlabeled channel estimates. Our self-supervising method learns general-purpose channel features robust to fading and system impairments. Learned representations are easily transferable to new environments and ready to use for other wireless downstream tasks. To the best of our knowledge, the proposed method is the first joint-embedding self-supervised approach to forsake the dependency on contrastive channel estimates. Our approach outperforms fully-supervised techniques in small data regimes under fine-tuning and, in some cases, linear evaluation. We assess the performance in centralized and distributed massive MIMO systems for multiple datasets. Moreover, our method works indoors and outdoors without additional assumptions or design changes.
	\end{abstract}
	
	\section{Introduction}
	Precise radio localization has been a long-standing research challenge. Presently, the most notable methods of radio localization are \textit{space-based}, specifically the global navigation satellite systems, such as global positioning systems (GPS). Today, a multitude of commercial and non-commercial applications rely on GPS \cite{kaplan2017understanding}. GPS-enabled mobile equipment mainly depends on the time-of-arrival (TOA) information of the received signal to evaluate the range to multiple synchronized satellite landmarks; under favorable conditions this allows for a localization accuracy of a few meters or even below. However, the GPS quality of service (QoS) is significantly impaired in dense urban and indoor areas, where the environment is characterized by multipath effects and additional pathloss due to signal blockages. Therefore, endeavors to improve the localization services have also been sustained as part of \textit{ground-based} wireless communication systems. The most distinguished approaches among the \textit{ground-based} systems have undergone advancements in cellular standards \cite{del2017survey}. Similar to GPS, the latest standardized wireless communication systems use dedicated pilot resources to measure the signal information at multiple synchronized base stations (BS) to determine the position of the user equipment (UE). Some strategies involve measuring the received signal strength (RSS), angle-of-arrival (AOA), time-of-arrival, or a combination of them \cite{8682509,zane2020performance,wen2019survey}.
	
	Despite the advancements in localization research and significant performance improvements over the decades, emerging location-aware applications persistently challenge the state-of-the-work. Services with meter-level accuracy and very-high availability requirements are anticipated to be supported in fifth-generation (5G) and beyond communication systems \cite{3gpp.22.261}. New use cases impose additional strict requirements regarding accuracy, availability, and coverage. For instance, augmented reality relating services would require coverage of less than $10$ meters, an accuracy of a few centimeters, and a latency below $20$ ms. On the other hand, remote monitoring, asset tracking, and other internet of things related services require less than $10$ m accuracy and more than $1$ km coverage \cite{3gpp.22.261}.
	
	Furthermore, most use cases require a guaranteed QoS with availability ranging from $80\%$ to $99.9\%$ \cite{3gpp.22.261}. Cellular localization for the next-generation mobile communications systems is the most promising candidate to complement GPS to attain the QoS metrics of varying applications and use cases. Since communication, positioning, and sensing are expected to co-exist, new generation of mobile communications systems is additionally anticipated to leverage location information for self-optimization to improve efficiency and sustainability. 
	
	Current deployment of massive multiple-input multiple-output (MIMO) technology, a foreseen adoption of reconfigurable intelligent surfaces (RIS), and tenacious exploration of higher bandwidths in millimeter waves and beyond aid to improve angular- and time-granularity of signal information \cite{wymeersch2022radio}. This facilitates closing the gap between the aforementioned QoS requirements of different use cases and the performance provided by wireless communication and localization. However, such advancements are subject to increased computational complexity and scalability issues. To address such issues, a paradigm shift from classical to learning-based signal processing methods has been widely considered over the past years. Most notably, deep neural network (DNN) models have been proposed to benefit the overall next generation of communications systems for enhanced physical layer (PHY), improved efficiency, and simplified deployment. Proposed DNN models have been reported beneficial for a variety of fundamental wireless communications tasks, including, but not limited to, channel estimation, channel quantization, beamforming, and localization \cite{elijah2022intelligent, schwarz2021codebook,9616218}.
	
	Whereas a common approach in cellular-based localization systems is to rely on time and angular estimation of the signal from multiple base stations, such approaches generally are not limited to line-of-sight (LOS) but are much less accurate when LOS is unavailable. On the other hand, DNN models use CSI at a massive MIMO base station to train a supervised model with known channel prints and determine unknown transmitter's position during the operation phase.
	
	Predominantly, previous works use raw CSI to feed their respective proposed DNN architectures. However, models that rely solely on standard feedforward multi-layer perceptrons (MLP) cannot learn sufficiently high-quality representations to cope with imperfect channel estimates, temporal variations in the environment, and other system impairments. Hence, various works have addressed the issue by suggesting to hand-design more robust features, mainly by exploiting the approximately sparse angle- and delay-domain channel representation in a MIMO-OFDM system. For example, \cite{9348191} suggests utilizing a decimated delay-domain CSI representation followed by autocorrelation to capture system-invariant features. Yet, hand-crafting the input constrains the set of DNNs functions, accordingly limiting the achievable quality of learned representations.
	
	Similarly, numerous works suggest utilizing convolutional neural networks (CNNs) to better capture the channel characteristics relevant for the channel-to-location mapping \cite{vieira2017deep,de2020mamimo,ayyalasomayajula2020deep}. However, while CNNs are primarily designed and well-suited for structured signals (e.g., images), they introduce a very strong inductive bias by utilizing fix-sized filters to \textit{slice} the channel and learn different portions of the input channel representation. Although there are no rigorous attempts to justify its benefits in wireless localization, the inductive bias in CNNs should help learn wireless channel features in relatively static environments where it is relatively easy to isolate the stationary regions of the channel, and CNN-inherent assumption about translation equivariance can hold. This is helpful, especially in a small data regime, where CNNs can learn faster with less number of parameters. Nevertheless, CNNs inevitably restrict the model's ability where there are abundant training examples. Furthermore, translation equivariance is not necessarily a shared property between images and communication signals.
	Finally, applying convolutional kernels limit CNNs to relying solely on short-term dependencies obtainable within its \textit{receptive field}. Approaches that follow \cite{he2016deep} can drastically improve the respective \textit{receptive field}. However, they do so by fusing (e.g., pooling) locally isolated channel features.
	
	In contrast to CNN-based models, transformer-based architectures in natural language processing (NLP) \cite{vaswani2017attention}, computer vision \cite{dosovitskiy2020image}, or wireless communications \cite{9833994}, adopt the self-attention mechanism \cite{vaswani2017attention}, imposing much less inductive bias. Consequently, they have greater learning capacity in big-data regimes \cite{zhao2020exploring}. For instance, wireless transformer (WiT) proposed in \cite{9833994} shows exceptional localization accuracy even in scenarios characterized by strong multipath propagation. Moreover, self-attention in WiT can capture the dependencies between nearby as well as distant subcarriers, beyond the \textit{receptive field} of existing CNN-based localization techniques. Lastly, by comparison to CNNs, transformers are less computationally costly and have faster training and inference times \cite{dosovitskiy2020image}. We note that historically, the attention mechanism was used on top of convolutional feature maps, which is reflected in recent growing literature in wireless localization \cite{10018917}, or even a combination with long short-term memory (LSTM) \cite{ruan2022hi}. However, our goal is to maintain the convolution-free property of WiT, and rather focus on the input representation of the channel. Therefore, we rely on transformer architecture and WiT as the fundamental building block and learning framework. Finally, in this work, we peek inside WiT and stress its capabilities with respect to various factors. 
	\subsection{Self-supervised Learning}
	Despite the astonishing achievable results and the above-stated advantages of supervised learning in general and transformer-based models in particular, they are highly sensitive to the small-data regime. Although there is ample CSI available at the base station, geo-tagged CSI samples are very scarce. Furthermore, the existing approaches train and evaluate the models for specific environments and tasks separately. Moreover, the transferability and adaptability of the pre-trained fully-supervised representations of the existing models are unknown. To address these fundamental concerns, we propose a self-supervised wireless channel representation learning method, referred to as SWiT, short for self-supervised wireless transformer. Although self-supervised learning (SSL) was discussed decades ago in \cite{bromley1993signature}, it has only recently drawn immense attention for its data efficiency and generalization capabilities. SSL utilizes unlabeled data as a form of supervision, and the learned representations can be used for a variety of downstream tasks. It has been the subject of significant research in the field of NLP \cite{devlin2018bert}, speech recognition \cite{baevski2020wav2vec} and is an active research area for computer visual representation learning \cite{grill2020bootstrap, bardes2021vicreg, 9709990}.
	
	Unlike prior works in wireless communications, we leverage the redundant and complementary information across different subcarriers to learn to predict the channel from a single input realization. Hence, in contrast to the mining of triplets discussed in \cite{salihu2020low}, and forcing to discriminate \textit{distinct} channels, we avoid the need to sample negative pairs and, in this work, the need to use a contrastive loss as in \cite{9709990}. The latter has multiple benefits concerning computational cost and, more importantly, generalization capabilities. For example, techniques that rely on mining the channel realizations based on the time-of-arrival, or time-stamp as in \cite{ferrand2021triplet}, are highly impacted by the non-stationary nature of the channel as well as the ability to estimate the time-of-arrival sequence accurately. Furthermore, we propose to incorporate a subcarrier-level \textit{pretext} task for wireless channel representation learning, which can be crucial to leverage the microscopic fading characteristics of the channel effectively. In essence, this principle is similar to the so-called \textit{region-level} pre-training tasks in computer vision demonstrated in previous studies such as \cite{doersch2015unsupervised} and more recent studies in conjunction with transformer \cite{li2021efficient, 9878641} or CNN \textit{backbones} \cite{bardes2022vicregl}.
	\subsection{Our Contributions}
	In this work, we propose a new framework for self-supervised learning of wireless channel representations. Our contributions in this paper can be further summarized as follows.
	\begin{itemize}
		\item We learn general-purpose channel representations solely based on the input channel estimates.
		\item To the best of our knowledge, the proposed method is the first self-supervised \textit{joint embedding} learning approach to forsake the dependency on contrastive examples for wireless channel representation learning.
		\item We split the learning framework into capturing microscopic and macroscopic characteristics of fading channel.
		\item We propose channel transformations beneficial to improving the robustness to fading and other system impairments of learned embeddings.
		\item We show that our self-supervised approach can outperform fully-supervised wireless localization techniques in small data regimes, sometimes even with a linear model with negligible computational complexity.
		\item The self-supervised representations are easily transferable to new environments and ready to use for other wireless downstream tasks.
		\item We thoroughly investigate the localization performance of a pure transformer-based architecture.
	\end{itemize}
	
	We structure the rest of this article as follows. In Section \ref{systemModel} we outline the general system model. In Section \ref{sec:swit} we provide background on the wireless transformer and present the proposed self-supervised approach. Then, in Section \ref{sec:experimentsAndResults}, we evaluate the performance and demonstrate the transferability of the proposed method. Furthermore, we report additional studies to enhance our understanding on the performance gains. Finally, in Section \ref{sec:conclusion}, we draw our conclusions.
	
	\textit{Notation:} A matrix is denoted by $\mathbf{Y}$, a vector by $\mathbf{y}$ and a scalar by $y$. The element of a matrix $\mathbf{Y}$ is $y_{m,n}$ and the Euclidean norm of vectors is $\| \cdot \|$. The cardinality of a set as well as the absolute value of a scalar are $| \cdot |$. The notations $\otimes$ and $\mathbb{E}\{\cdot\}$ denote the Kronecker product and the expectation of a random variable.
	
	\begin{figure}[!t] 
		\centering
		{%
			\includegraphics[width=0.40\linewidth]{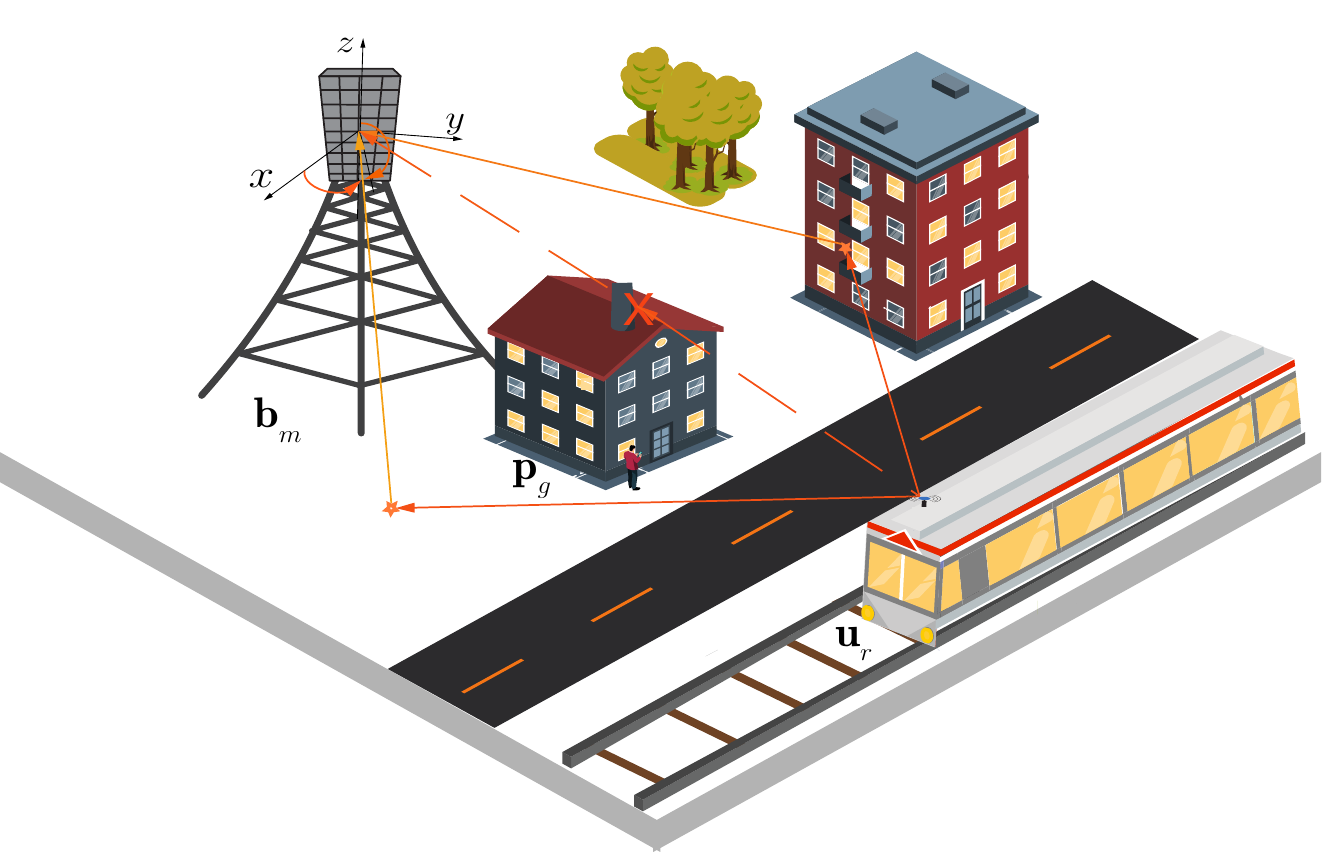}}
		\cprotect\caption{Illustration of the system model for co-located massive MIMO-OFDM scenario.} 
		\label{fig:System_model_SSL2022} 
	\end{figure} 
	\section{System Model and Problem Formulation}\label{systemModel}
	We consider an uplink transmission in a massive MIMO communication system, and assume $R$ single-antenna user locations placed at positions $\mathbf{u}_{r} = [u_{r,1},u_{r,2},u_{r,3}]^{T}$, where $r \in \{1, \ldots, R\}$. The base station (BS) is equipped with $N_{r}$ antenna elements. To keep this work more general, we also consider $N_{r}$ spatially distributed antennas among $M$ remote radio heads (RRHs) at positions $\mathbf{b}_{m} = [b_{m,1},b_{m,2},b_{m,3}]^{T} $, where $m \in \{1,\ldots,M\}$. Moreover, we consider $G$ scattering objects in the ROI at respective positions, $\mathbf{p}_{g} = [p_{g,1},p_{g,2},p_{g,3}]^{T}$ with $g \in \{1,\ldots, G\}$. 
	
	We consider OFDM-based transmissions, such that the channel transfer function used for localization is sampled over time (OFDM symbols) and frequency (OFDM subcarriers). The uplink input-output relationship of a single-user transmission from $\mathbf{u}_{r}$ on subcarrier $n\in\{1,\ldots,N_c\}$ is 
	\begin{equation}
		\mathbf{y}_{n} = \sqrt{P}\mathbf{\tilde{h}}_{n} x_{n} + \mathbf{n}_{n} \, . 
	\end{equation}
	Here, $P$ is the average transmit power, $x_{n}$ is the normalized transmitted signal with $|{x}_{n}|^2=1$, $\mathbf{\tilde{h}}_{n} \in\mathbb{C}^{N_r \times 1}$ is the channel vector, and $\mathbf{n}_{n} \sim \mathcal{C N}\left(\mathbf{0}, N_{0} \mathbf{I}_{N_r}\right)$, where $N_{0}$ is the noise power spectral density. We assume that the BS is able to estimate the channel vectors $\mathbf{\tilde{h}}_{n}$ through uplink pilot signals, such as in a system operating in time division duplex (TDD) mode, and can mitigate pilot contamination \cite{larsson2014massive, jose2011pilot}. We denote the estimated CSI as $\mathbf{{h}}_n$. Hence, the problem of location estimation in general is to derive an optimal accuracy of the unknown target position $\mathbf{u}_{r}$ from the estimated $\{\mathbf{{h}}_{n}\}_{\forall n}$.
	
	\subsection{Channel Model}\label{sec:ChannelModel}
	Our proposed DNN methods are data-based and, therefore, do not rely on a specific channel model. Nevertheless, to clarify the general relationship between the user location and the estimated parameters of the channel, we utilize a commonly adopted geometric channel model to characterize the channel for such scenarios as that illustrated in Fig. \ref{fig:System_model_SSL2022} \cite{heath2016overview},
	\begin{equation}
		\mathbf{\tilde{h}}_{n} = \sum_{g=1}^{G} \eta_{g} e^{j{2 \pi k}{\Delta f} \tau_{g}} \mathbf{a}\left(\varphi_{a z,g}, \varphi_{e l,g} \right) \, . 
	\end{equation}
	Here, $\eta_{g}$ and $\tau_{g}$ denote the $g-$th path's complex gain and propagation delay (ToA) between $r-$th user location and $m-$th BS. The angles of arrival (AoA) in azimuth and elevation are denoted by $\varphi_{a z,g}$ and $\varphi_{e l,g}$, respectively. The subcarrier spacing is denoted by $\Delta f $. Although it is challenging to argue the resolvability of multiple paths in lower frequencies, and hence their relationship to the geometry of the environment, in mmWave and beyond, there is a stronger relation between the individual received paths and the geometric information of the channel \cite{wymeersch2022radio, 7426565}. Finally, assuming a uniform planar array with antenna elements along $x-$ and $z-$axis, the expression for the array response vector for the elaborated example at the receiver can be written as $\mathbf{a}\left(\varphi_{\mathrm{az}}, \varphi_{\mathrm{el}}\right)=\mathbf{a}_{z}\left(\varphi_{\mathrm{el}}\right) \otimes \mathbf{a}_{x}\left(\varphi_{\mathrm{az}}, \varphi_{\mathrm{el}}\right)$.
	The array steering vectors $\mathbf{a}_{x}(\cdot)$, and $\mathbf{a}_{z}(\cdot)$ are 
	\begin{equation}
		\begin{aligned}
			\mathbf{a}_{x}\left(\varphi_{\mathrm{az}}, \varphi_{\mathrm{el}}\right)=\left[1, e^{j \frac{2\pi}{\lambda_c} d \sin \left(\varphi_{\mathrm{el}}\right) \sin \left(\varphi_{\mathrm{az}}\right)}, \ldots, e^{j \frac{2\pi}{\lambda_c} d\left(M_{x}-1\right) \sin \left(\varphi_{\mathrm{el}}\right) \sin \left(\varphi_{\mathrm{az}}\right)}\right]^{T},
		\end{aligned}
	\end{equation}
	\begin{gather*}
		\mathbf{a}_{z}\left(\varphi_{\mathrm{el}}\right)=\left[1, e^{j \frac{2\pi}{\lambda_c} d \cos \left(\varphi_{\mathrm{el}}\right)}, \ldots, e^{j \frac{2\pi}{\lambda_c} d\left(M_{z}-1\right) \cos \left(\varphi_{\mathrm{el}}\right)}\right]^{T}, 
	\end{gather*}
	with $\lambda_{c} = c/f_{c}$, where $f_{c}$ being the carrier frequency and $c$ the speed of light, and $d = \lambda_{c}/2$ the antenna element spacing.
	
	\subsection{Uncertainty Model}
	When generating synthetic channel data, we consider scenarios in which a subset of scattering objects vary its position over time $t \in \{1, \ldots, T \}$  (e.g., moving objects in the environment), and assume ${p}_{g,i}^{(t)} \triangleq {p}_{g,i} + \mathring{w}_{g,i}$, where $\mathring{w}_{g,i}$ is the zero-mean Gaussian noise term with variance ${\sigma}_{\mathring{w}}^{2}$ at $i-$th coordinate. Similarly, we account for an offset in the position of antennas of the transmitter, ${u}_{r,i}^{(t)} \triangleq {u}_{r,i} + \dot{w}_{r,i}^{(t)}$ where $\dot{w}_{g,i}$ denotes the zero-mean Gaussian noise with variance ${\sigma}_{\dot{w}}^{2}$ at $i-$th coordinate. We note that the variations in the position of the scatterers alter the gain, delay and angle information of the individual multi-bounce non-line-of-sight (NLOS) paths. The uncertainty in the position of UE antenna, allows us to account for the effect of imperfect channel estimates due to e.g., lack of perfect synchronization. 
	
	Finally, stacking $N_c^{\prime} \leq N_{c}$ subcarriers in a matrix form for the $r-$th user location leads to
	\begin{equation}
		\mathbf{H}_{r}^{(t)} =\left[\mathbf{{h}}_{1}^{(t)}, \mathbf{{h}}_{2}^{(t)}, \ldots, \mathbf{{h}}_{N_{c}'}^{(t)}\right] \in \mathbb{C}^{N_{r} \times N_{c}^{\prime}} ,
	\end{equation}
	where $N_c'$ may correspond to pilot subcarriers.
	For convenience, in the following we drop the superscript $t$ from our notation.
	
	\subsection{Problem Formulation}
	A fundamental problem in deep wireless communications is to learn compressed representations of the channel that nonetheless can be used to efficiently realize varying downstream tasks, such as wireless localization. Thus, our goal in this work is to learn a channel representation $\mathbf{\bar{o}}_{r}$ (i.e., an embedding), using neither labels nor a contrastive objective function. To formalize deep wireless SSL, we first construct a \textit{backbone} neural network with a dual encoder-projector framework, which shares the same overall learning paradigm with \cite{grill2020bootstrap, 9709990}. The encoders, denoted by $f_{\boldsymbol{\Theta}}(\cdot)$ and $f_{\boldsymbol{\Psi}}(\cdot)$, and the projectors denoted by $g'_{\boldsymbol{\Theta}}(\cdot)$ and $g'_{\boldsymbol{\Psi}}(\cdot)$ are parameterized by $\boldsymbol{\Theta}$, and $\boldsymbol{\Psi}$, respectively. Given two input representations of the same channel  realization, $\mathbf{{H}}_{r}$ and $\mathbf{{H}}_{r}^{(+)}$ (e.g., two augmented views), we first write a common SLL formulation as 
	\begin{equation}
		\begin{aligned}
			\left\{	\boldsymbol{\Theta}^{\star}, \boldsymbol{\Psi}^{\star} \right\}  = \argmin_{\boldsymbol{\boldsymbol{\Theta}}, \boldsymbol{\Psi}} \mathcal{L}_{\text{SSL}}\left(g'_{\boldsymbol{\Theta}}(f_{\boldsymbol{\Theta}}(\mathbf{{H}}_r)), g'_{\boldsymbol{\Psi}}(f_{\boldsymbol{\Psi}}(\mathbf{{H}}_r^{(+)}))\right).
		\end{aligned}
	\end{equation}
	where $\mathcal{L}_{\text{SSL}}(\cdot)$ is the SSL loss function, which we will detail in Sec. \ref{sec:swit}. Having a set of optimal parameters $\boldsymbol{\Psi}^{\star}$, our goal then becomes to fine-tune the parameters of the encoder $f_{\boldsymbol{\Psi^{\star}}}(\cdot)$ (a.k.a. momentum encoder of the \textit{backbone} network) altogether with a linear multi-layer perception (MLP) head $g_{\boldsymbol{\Phi}}(\cdot)$ to learn a mapping function between the channel $\mathbf{H}_r$ and user location information, $\mathbf{u}_{r}$,
	\begin{equation}
		\begin{aligned}
			\left\{ \boldsymbol{\Psi}^{\star\star}, \boldsymbol{\Phi}^{\star} \right\}
			&= \argmin_{\boldsymbol{\Psi^{\star}}, \boldsymbol{\Phi}} \mathcal{L}_{\text{SUP}}\left(\mathbf{u}_{r},g_{\boldsymbol{\Phi}}(f_{\boldsymbol{\Psi}^{\star}}(\mathbf{H}_r))\right) \\
			&= \argmin_{\boldsymbol{\Psi^{\star}}, \boldsymbol{\Phi}} \, \mathbb{E}\left[\left\| \mathbf{u}_{r} - g_{\boldsymbol{\Phi}}(f_{\boldsymbol{\Psi}^{\star}}(\mathbf{H}_r)) \right\|^{2}\right].
		\end{aligned}
	\end{equation}
	Alternatively, in case of linear evaluation, we aim to only train the linear MLP head on top of the \textit{frozen} parameters of the encoder $f_{\boldsymbol{\Psi^{\star}}}(\cdot)$, and only update $\boldsymbol{\Phi}$, i.e., 
	\begin{equation}
		\begin{aligned}
			\boldsymbol{\Phi}^{\star}
			&= \argmin_{\boldsymbol{\Phi}} \mathcal{L}_{\text{SUP}}\left(\mathbf{u}_{r},g_{\boldsymbol{\Phi}}(f_{\boldsymbol{\Psi}^{\star}}(\mathbf{H}_r))\right)\,. 
		\end{aligned}
	\end{equation}
	\section{SWIT: Self-supervised Wireless Transformer}\label{sec:swit}
	Since we extensively utilize a transformer model in our learning framework, in Sec. \ref{sec:Preliminaries} we first review the main building blocks of the wireless transformer model introduced in \cite{9833994}. Subsequently, we elaborate on our self-supervised channel representation learning framework.
	
	\subsection{Preliminaries}\label{sec:Preliminaries}
	\begin{figure}[!t] 
		\centering
		{%
			\includegraphics[width=0.45\linewidth]{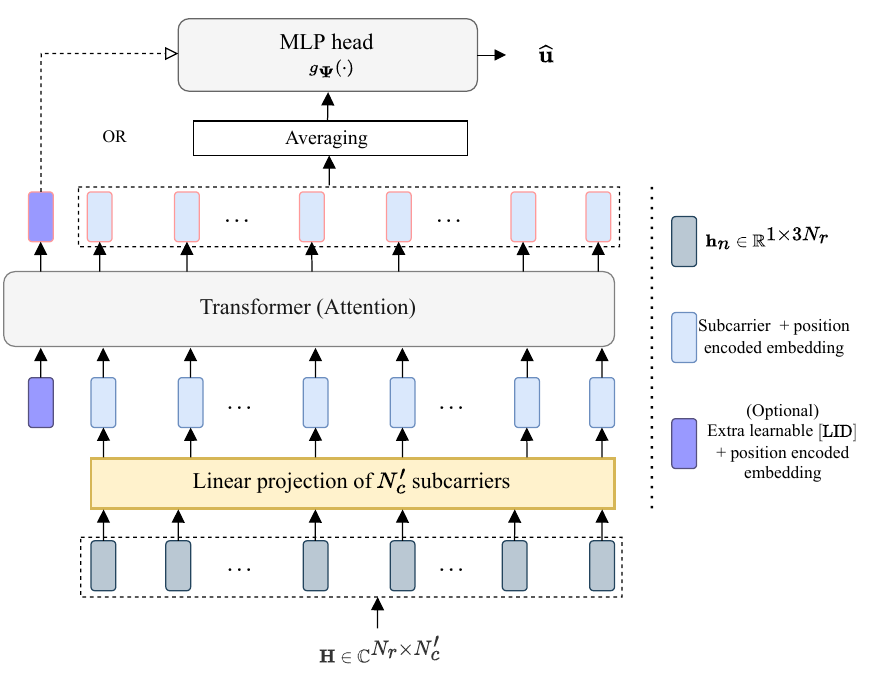}}
		\cprotect\caption{Overview of WiT \cite{9833994}. Linearly map each subcarrier into embeddings, add positional encodings, and feed the feature vectors to a deep transformer.} 
		\label{fig:WiT_architecture_overview} 
	\end{figure} 
	
	\subsubsection{Wireless Transformer} As detailed in \cite{9833994}, and depicted in Fig. \ref{fig:WiT_architecture_overview}, the input channel $\mathbf{{H}}_r \in \mathbb{C}^{N_r \times N_c^{\prime}}$ is viewed as a sequence of subcarriers $\{\mathbf{h}_{n} \in \mathbb{R}^{1\times3N_r} \}_{n=1}^{N_c'}$ represented as the real, imaginary, and absolute parts of the row channel vector. Then, each of the subcarrier representations is linearly embedded. Further, positional encodings are added, and the subcarrier representations are fed to a transformer block with an attention module for feature extraction. For the MLP head, one can either average over the attended features, or use an extra learnable symbol, denoted by $\left[ \verb*|LID| \right] $. In the sequel, we further describe the signal processing through a single transformer block, which is depicted in Fig. \ref{fig:attention_module}.
	\subsubsection{Linear Embeddings} Each subcarrier is linearly transformed into an embedding through a linear layer with learnable parameters $\mathbf{E} \in \mathbb{R}^{3N_{r} \times D}$, i.e., $\mathbf{e}_{i} = \mathbf{h}_{i}\mathbf{E}$.
	Since transformer-based models use neither recurrence nor a standard convolutional operator, the attention module treats all the subcarrier representations as permutation invariant. Therefore, it lacks the possibility to exploit the sequence arrangement of frequency-dependent subcarriers. Thus, we encode positional information to learn to represent the subcarrier's position in the sequence.
	\subsubsection{Subcarrier Positional Encoding}
	We assign a learnable real-valued vector embedding $\mathbf{g}_{i}\in \mathbb{R}^{1\times D}$ to each subcarrier index $i$. Then, given the input channel, $\mathbf{{g}}_{i}$ is added to the subcarrier embedding $\mathbf{e}_{i}$ at position $i$. Hence, the input to the transformer block becomes $\mathbf{\hat{e}}_i= \mathbf{e}_{i} + \mathbf{{g}}_{i}$.
	\subsubsection{Location Identification}
\begin{figure}[!t]
		\centering
		\begin{minipage}{.32\linewidth}
			\centering
			\begin{subfigure}[t]{0.6\linewidth}
				\includegraphics[width=\textwidth]{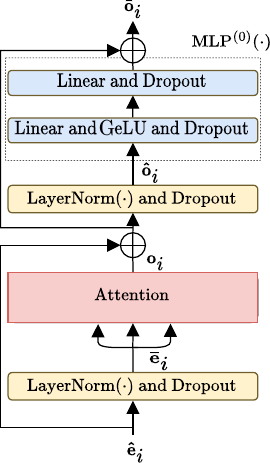}
				\caption{Transformer block.}
				\label{fig:attention_module}
			\end{subfigure}
		\end{minipage}
		\begin{minipage}{.30\linewidth}
			\centering
			\begin{subfigure}[t]{0.6\linewidth}
				\includegraphics[width=\textwidth]{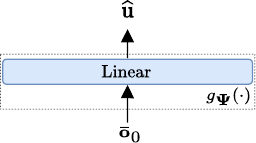}
				\caption{MLP head.}
				\label{fig:mlp_head}
			\end{subfigure}
		\end{minipage}
		\begin{minipage}{.30\linewidth}
			\centering
			\begin{subfigure}[t]{0.6\linewidth}
				\includegraphics[width=\textwidth]{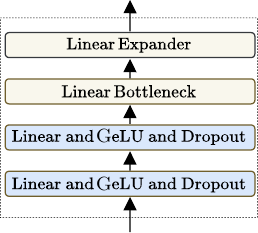}
				\caption{Projectors.}
				\label{fig:base_dnn}
			\end{subfigure} 
		\end{minipage}
		\caption{DNN model details. a) Transformer block with the attention module, b) Linear MLP-head for location estimation, and c) shows the global and local projector DNNs useful for SWiT.}
		\label{fig:dnn_model_details}
	\end{figure}

	To add \textit{global} context information on the whole channel, we prepend to the set of subcarrier embeddings a special symbol $\left[ \verb*|LID| \right]$. This has an analogous role to $\left[ \verb*|CLS| \right]$ token introduced in original BERT transformer \cite{devlin2018bert}. The special symbol $\left[ \verb*|LID| \right]$ is a learnable vector, denoted as $\mathbf{\hat{e}}_{0} \in \mathbb{R}^{D}$, whose representation is a compressed characterization of the whole channel from the $r-$th user. We use its final representation (i.e., $\mathbf{\bar{o}}_{0} \in \mathbb{R}^{D}$) to feed into a fully connected linear layer detailed in Fig. \ref{fig:mlp_head} for the final mapping of features to location. We will also utilize the special vector for self-supervision and transfer-learning investigation. The size of the set of input vectors to the transformer block is $C = N_{c}' + 1$.
	\subsubsection{Attention}
	In case of self-attention \cite{vaswani2017attention}, we consider three input \textit{copies} and project them using the same set of weights, $\mathbf{{W}}_{q} = \mathbf{{W}}_{k} = \mathbf{{W}}_{v}$. To this end, we write the self-attention as 
	\begin{equation}
		\begin{aligned}
			\mathbf{o}_{i} = \sum_{j=1}^{C} \frac{\exp \left(\alpha_{i, j}\right)}{\sum_{j^{\prime}=1}^{C} \exp \left(\alpha_{i,j^{\prime}}\right)}\left(\mathbf{\bar{e}}_{j} \mathbf{{W}}_{v} \right) \\[6pt]
		\end{aligned} \label{eqn:AttendedEmbedding} 
	\end{equation}
	where $\alpha_{i,j}$ is the attention coefficient between the two embeddings at positions $i$ and $j$, 
	\begin{equation}
		\begin{aligned}
			\alpha_{i,j} = \frac{1}{\sqrt{D}}\left(\mathbf{\bar{e}}_{i} \mathbf{{W}}_{q} \right)\left(\mathbf{\bar{e}}_{j} \mathbf{{W}}_{k} \right)^{T} .
		\end{aligned} 
	\end{equation}
	In above, $\mathbf{\bar{e}}_{i} = \mathrm{LayerNorm}(\mathbf{\hat{e}}_{i}; \zeta, \iota) $ where $\zeta$ and $\iota$ are hyperparameters \cite{ba2016layer}.
	\subsection{SWiT}
\begin{figure*}[t]%
		\centering 
		\includegraphics[width=0.75\linewidth]{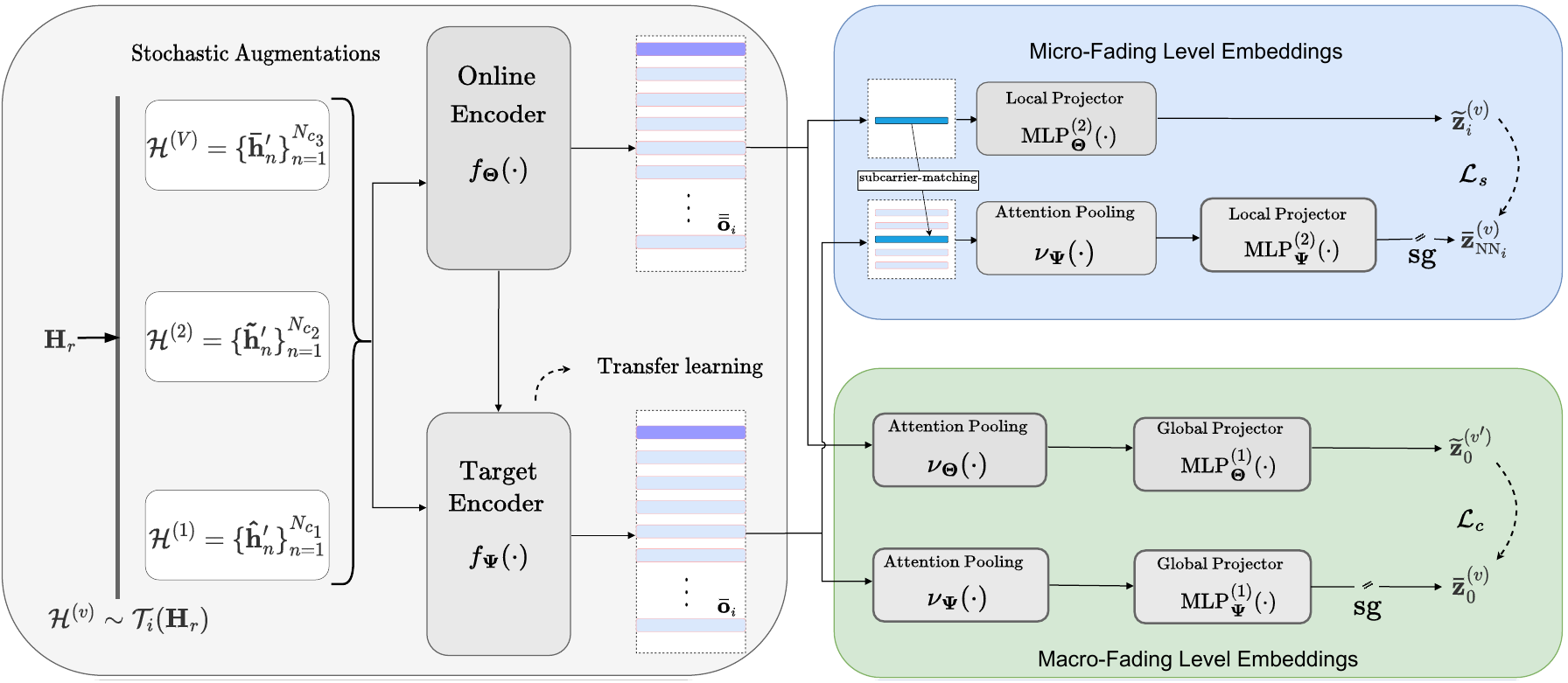}%
		\caption{Both online and target models take multiple views of the same input channel realization. Target encoder serves for transfer learning.}
		\label{fig:SWiTArchitectureAll}
	\end{figure*}

	We propose SWiT, a new method for self-supervised wireless channel representation learning. We aim to learn channel representations invariant to perturbations modeled as augmentations from a single channel realization. As depicted in Fig. \ref{fig:SWiTArchitectureAll}, SWiT is a two-branch deep neural network, which we denote as online and target networks, respectively. These two networks, in most of the parts, share the same architectural design. However, they have different sets of weights. First, we define a set of stochastic transformations $\mathcal{Q} := \{Q_1, \dots, Q_{A}\}$, and associate a probability of occurence for each. Furthermore, we define $\mathcal{T}_{i} (\mathbf{H_r}) = \left(Q_{A} \circ Q_{A-1} \circ ...Q_{1}\right) (\mathbf{H}_r)$. Then, the role of the stochastic augmentation module is to output two views of the channel comprising of most of the subcarriers, i.e., 
	\begin{equation}
		\begin{aligned}
			\{\mathbf{\bar{h}}_{n}^{\prime} \in \mathbb{R}^{3N_r}\}_{n=1}^{N_{c_1}} 
			& \triangleq \mathcal{T}_{1}\left(\mathbf{H_r}\right)
		\end{aligned} \label{eqn:globalView1_transf}  
	\end{equation}
	for the first view, and similarly for the second view,
	\begin{equation}
		\begin{aligned}
			\{\mathbf{\tilde{h}}_{n}^{\prime} \in \mathbb{R}^{3N_r}\}_{n=1}^{N_{c_2}} & \triangleq \mathcal{T}_{2}\left(\mathbf{H_r}\right) \, ,
		\end{aligned} \label{eqn:globalView2_transf} 
	\end{equation}
	where $N_{c_1}$ and $N_{c_2}$ denote the total number of selected subcarriers for the respective transformation. We will refer to these transformations as \textit{global} views, and we consider $N_{c_2} \leq N_{c_1}$. Similarly, the augmentation module also outputs $N_s$ random channel views,
	\begin{equation}
		\begin{aligned}
			\{\mathbf{\hat{h}}_{n}^{\prime} \in \mathbb{R}^{3N_r} \}_{n=1}^{N_{c_3}} & \triangleq \mathcal{T}_{3}\left(\mathbf{H_r}\right) \, ,
		\end{aligned} \label{eqn:localView_transf} 
	\end{equation}
	where $N_{c_3} \ll N_{c_2}$, and we refer to them as \textit{local} views. Hence, in total we have $V = 2+N_s$ views of the same input channel.
	Both \textit{global} and \textit{local} channel views are sent consecutively through the online $f_{\boldsymbol{\Theta}}(\cdot)$ and target encoder $f_{\boldsymbol{\Psi}}(\cdot)$. Consequently, producing the corresponding representations for a single view,
	\begin{equation}
		\begin{aligned}
			\{\mathbf{\bar{o}}_n \in \mathbb{R}^{D}\}_{n=1}^{C} := f_{\boldsymbol{\Psi}} \left(	\{\mathbf{\bar{h}_{n}^{\prime}}\}_{n=1}^{N_{c_1}}\right) \, ,
		\end{aligned} 
	\end{equation}
	and similarly for the others, which are further detailed in Sec. \ref{sec:macroFadingLevel}.
	After obtaining the representations from the encoder, we split the learning into two separate projectors, namely, micro-fading and macro-fading channel representation learning projectors. By doing so, we aim to better capture the large-scale channel characteristics along with small-scale, per-subcarrier features. The loss is computed on the final embeddings of views obtained from online and target encoders.
	Before describing the large-scale and small-scale feature learning modules shown in Fig. \ref{fig:SWiTArchitectureAll}, we first elaborate the set of designed augmentations. 
	
	\subsection{Stochastic Channel Augmentations}\label{sec:channelAugmentations}
	Learned representations should remain invariant to their corresponding views or channel transformations. This implies that the design of augmentations, or view-selection functions $\mathcal{T}_{i}(\cdot)$, is an essential factor for the influence of learned invariances that yield useful channel representations. As we already mentioned, the augmentation module's role is to transform an input channel realization into a pre-defined number of correlated views. Specifically, we sequentially apply the following channel transformations. We first apply \textit{random subcarrier selection (RSS)} followed by resizing of the grouped subcarriers to a fixed size, and optionally apply \textit{random subcarrier flipping (RSF)}. Furthermore, we apply a \textit{random gain offset (RGO)}, followed by a \textit{random fading component (RFC)}, \textit{random sign change (RSC)}, \textit{normalization}, and optionally we add \textit{Gaussian noise}.
	
	\subsubsection{Random Subcarrier Selection (RSS)} 
	We select $N_{c_i}' = \lfloor \gamma_{i} {N}_c' \rfloor$ adjacent subcarriers and, for practical reasons, linearly interpolate to resize the whole channel to a fixed size $N_r \times \widetilde{N}_c'$, i.e., $\widetilde{N}_c'$ subcarrier representations for the input to the encoders for \textit{global} views. Similarly, for the \textit{local} views, we set a fixed size of $N_r \times \bar{N}_c'$ before feeding to the online encoder. As already mentioned, we feed the network with two \textit{global} views and $N_s$ \textit{local} views, each comprising of a few subcarriers. Unless mentioned otherwise, we select $\gamma_{c_{1}} = 0.9$, and $\gamma_{c_{2}} = 0.8$, for the \textit{global} views and $\gamma_{c_3}=0.1$ for the \textit{local} views. By doing so, we force the network to learn subcarrier-to-channel mappings. We also experimented with other subcarrier sampling strategies. For example, we considered selecting every third subcarrier, $N_{c}' = N_{{c_i}}' \quad(\bmod \; 3)$. However, during the experiments, we observed no significant gains.
	
	\subsubsection{Random Subcarrier Flipping (RSF)}
	We optionally apply mirroring, where each subcarrier representation is flipped as 
	\begin{equation}
		\begin{aligned}
			Q_a(n) = \widetilde{N}_c'-(n-1) \qquad n=1,\ldots, \widetilde{N}_c' \, ,
		\end{aligned} 
	\end{equation}
	for two \textit{global} views, and similarly for all other $N_s$ views.
	\subsubsection{Random Gain Offset (RGO)} Selecting a subset of subcarriers alone yields representations that may share the same channel gain distribution between antenna elements at any two subcarrier, enabling the model to exploit this property, and quickly minimize the loss. However, by doing so, the network may fail to capture generalizable channel characteristics. To circumvent such phenomenon, we scale instantaneous channel coefficients by a constant offset, $\xi_{\text{o}} = 1 \pm \xi_{\text{rgo}}$,
	\begin{equation}
		\begin{aligned}
			Q_a(h_{n,i}) = \xi_{\textrm{o}} {h_{n,i}} \qquad \forall i = 1, \ldots, 3N_r \, .
		\end{aligned} 
	\end{equation}
	Through experimentation, we found that ${\xi}_{\text{rgo}} \sim \mathcal{U}(0,0.1)$ is sufficient.
	\subsubsection{Random Fading Component (RFC)} To account for the robustness to additional multipath components, we shift the instantaneous peak power of the channel. We do so by applying a Rayleigh distribution only to the absolute part of the transformed view,
	\begin{equation}
		\begin{aligned}
			Q_a(h_{n,i}) = \frac{{h_{n,i}}}{\sigma_{\text{rfc}}^{2}} \exp \left(-{h_{n,i}^2} /\left(2 \sigma_{\text{rfc}}^2\right)\right) \quad \forall i = 2N_r, \ldots, 3N_r \, .
		\end{aligned} 
	\end{equation}
	Here, we find the scale factor $\sigma_{\text{rfc}} \sim \mathcal{U}(0.5, 0.6)$ through experimentation. Furthermore, in order to help the model in improving its ability to handle uncertainty, we apply \textit{RFC} only to the second view, denoted as $\mathcal{H}^{(2)}$ in Fig. \ref{fig:SWiTArchitectureAll}. 
	\subsubsection{Random Sign Change (RSC)} We randomly negate all real-valued channel coefficients for the second view (i.e., $\mathcal{H}^{(2)}$),
	\begin{equation}
		\begin{aligned}
			Q_{a}(h_{n,i}) = (-1)h_{n,i} \qquad \forall i = 1, \ldots, 3N_r .
		\end{aligned} 
	\end{equation}
	This technique can be understood as a form of introducing an adversarial example, thereby increasing the robustness of the model.
	\subsubsection{Normalization} We finally post normalize all the channel transformations by dividing real, imaginary and magnitude parts with the corresponding maximum absolute value in it, $\Delta_{\operatorname{Re}} = \max(\max(\{|\mathbf{{H}}_{r}^{(\operatorname{Re})}|\}_{r=1}^{R}))$. Similarly, we normalize the imaginary, $\Delta_{\operatorname{Im}}$, as well as the absolute part, $\Delta_{\operatorname{Abs}}$.
	
	\subsubsection{Gaussian Noise} Also, we occasionally inject additional randomness by adding Gaussian noise to the normalized channel,
	\begin{equation}
		\begin{aligned}			
			Q_{a}(h_{n,i}) = h_{n,i} + {\omega}_{n,i} \, ,
		\end{aligned}
	\end{equation}
	where ${\omega}_{n,i}$ is the zero-mean Gaussian noise with variance $\mathbf{\sigma}_{Q}^{2}$.
	\subsection{Macro-fading Level Representations}\label{sec:macroFadingLevel}
	We train SWiT in a self-supervised manner by predicting the transformed views of the channel with varying characteristics from each other in the embedding space of the online and target networks. Specifically, the online network processes both \textit{local} as well as \textit{global} channel views (i.e., $V = 2+N_s$ views) to output the respective embeddings   
	\begin{equation}
		\begin{aligned}
			\mathbf{\widetilde{z}}_0^{(v)} = \frac{\exp (\mathbf{{\dot{z}}}_0^{(v)} \chi_{\boldsymbol{\Theta}}^{-1})}{\sum_{j=1}^n \exp ({\dot{z}}_{0,j}^{(v)} \chi_{\boldsymbol{\Theta}}^{-1})} \,,
		\end{aligned} \label{eqn:normalizLIDOnline}
	\end{equation}
	where $v$ has been added to make the index of the channel's views explicit. In (\ref{eqn:normalizLIDOnline}), $\chi_{\boldsymbol{\Theta}}$ is a temperature function that controls the peak of the output distribution for the online network.
	The embeddings ${\mathbf{\dot{z}}}_{0}^{(v)}$ are mapped from the global projector as 
	\begin{equation}
		\begin{aligned}
			{\mathbf{\dot{z}}}_{0}^{(v)} = \mathrm{MLP}_{\boldsymbol{\Theta}}^{(1)}  \circ \nu_{\boldsymbol{\Theta}} (\{\mathbf{\bar{\bar{o}}}_n^{(v)}\}_{\forall n})
		\end{aligned} \label{eqn:projectorGlobalOnline} \,,
	\end{equation}
	where the intermediate function, $\nu(\cdot)$, is denoted as attention pooling operation, implemented as a transformer block, and $\mathbf{\bar{\bar{o}}}_n^{(v)} \in \mathbb{R}^{D}$ is the output from the online encoder.
	
	On the other hand, for the target neural network branch, we evaluate 
	\begin{equation}
		\begin{aligned}
			\mathbf{\bar{z}}_0^{(v)} = \frac{\exp ((\mathbf{{\mathring{z}}}_0^{(v)} - \boldsymbol{\epsilon}_{0}) \chi_{\boldsymbol{\Psi}}^{-1})}{\sum_{j=1}^n \exp (({\mathring{z}}_{0,j}^{(v)} - {\epsilon}_{0,j}) \chi_{\boldsymbol{\Psi}}^{-1})}\,. 
		\end{aligned} \label{eqn:normalizLIDTarget}
	\end{equation}
	Likewise, $\chi_{\boldsymbol{\Psi}}$ is the temperature function for the target model that controls the peak of the output distribution. The $\boldsymbol{\epsilon}$ indicates the centering of the ouput distribution, and we later detail it in Sec. \ref{subsec:Optimization}. The embeddings ${\mathbf{\mathring{z}}}_{0}^{(v)}$ are mapped from the global projector of the target network as 
	\begin{equation}
		\begin{aligned}
			{\mathbf{\mathring{z}}}_{0}^{(v)} = \verb*|sg|(\mathrm{MLP}_{\boldsymbol{\Psi}}^{(1)}  \circ \nu_{\boldsymbol{\Psi}} (\{\mathbf{\bar{o}}_n^{(v)}\}_{\forall n}))
		\end{aligned} \label{eqn:projectorGlobalTarget} \,,
	\end{equation}
	where $\verb*|sg|$ is the $\verb*|stop-gradient|$ operation.
	
	To match the distribution between the embedding outputs of the target and online networks across different views, we compute the cross-entropy to compare each \textit{global} view from the target encoder with any other alternate view from the online encoder,
	\begin{equation}
		\begin{aligned}
			\mathcal{L}_{c} :=- \frac{1}{2(V-1)}\sum_{v=1}^{2}\sum_{v'\neq v}^{V}\bar{\mathbf{z}}_{0}^{(v)}\log \left(\tilde{\mathbf{z}}_{0}^{(v')}\right) \, .
		\end{aligned} \label{eqn:loss_globView} 
	\end{equation}
	
	\subsection{Micro-fading Level Representations}
	As mentioned earlier, in addition to capturing the macroscopic fading characteristics of the channel, our goal is to exploit microscopic fading characteristics too. To achieve this, we propose to learn subcarrier-level representations, as depicted in the upper-right part of Fig. \ref{fig:SWiTArchitectureAll}. Specifically, we assume neighboring subcarriers as positive examples. By doing so, i.e., performing a limited neighborhood search, we maintain lower computational complexity, as opposed to considering all subcarrier representations. Therefore, for each embedding $\bar{\mathbf{o}}_i^{(v)}$, we first evaluate its correlation with the representations in its neighborhood $\{\bar{\mathbf{o}}_j^{(v)}\}_{j\in \mathcal{N}_i}$, where $\mathcal{N}_i$ is the set of indices of adjoint subcarrier-representations, $|\mathcal{N}_i| = K_n$, and $K_n \ll N_c'$. We sort the indices based on the similarity values in the descending order,
	\begin{equation}
		\begin{aligned}
			\mathcal{N}'_i = \underset{\substack{K_n \\ j \in \mathcal{N}_i}}{\operatorname{sort} \max } \, \rho \left(\bar{\mathbf{o}}_{i}^{(v)}, \bar{\mathbf{o}}_{j}^{(v)}\right)
		\end{aligned} \label{eq:sort} \, ,
	\end{equation}
	where 
	\begin{equation}
		\rho(\bar{\mathbf{o}}_{i}^{(v)}, \bar{\mathbf{o}}_{j}^{(v)}) = \frac{\langle \bar{\mathbf{o}}_{i}^{(v)} \, , \bar{\mathbf{o}}_{j}^{(v)}  \rangle}{\left\|\bar{\mathbf{o}}_{i}^{(v)}\right\| \left\|\bar{\mathbf{o}}_{j}^{(v)}\right\|}
	\end{equation}
	is the measure of similarity, i.e., the correlation between $\bar{\mathbf{o}}_i^{(v)}$ and embeddings in its neighbourhood $\mathcal{N}_i$. Finally, we select $K_{k} = |\mathcal{P}_i|$, where $\mathcal{P}_i \subset \mathcal{N}'_i$, represents the $\verb*|top-k|$ matched indices. 
	
	Instead of fusing (e.g., averaging) $\verb*|top-k|$ matched subcarrier-representations, we pass $\{\bar{\mathbf{o}}_j^{(v)}\}_{j\in \mathcal{P}_i}$ to another transformer block, and output a compressed representation $\mathbf{z}_{\textrm{NN}_i}^{(v)}$, 
	\begin{equation}
		\mathbf{z}_{\textrm{NN}_i}^{(v)} := \nu_{\boldsymbol{\Psi}}(\{\bar{\mathbf{o}}_j^{(v)}\}_{j\in\mathcal{P}_i}) \,. 
	\end{equation}
	We feed the small-scale embeddings through online and target feedforward MLP networks $\mathrm{MLP}_{\boldsymbol{\Theta}}^{(2)}(\cdot)$ and $\mathrm{MLP}_{\boldsymbol{\Psi}}^{(2)}(\cdot)$, respectively. To this end, we write the embeddings $\mathbf{\dot{z}}_i^{(v)}$ and $\mathbf{\mathring{z}}_{\mathrm{NN}_i}^{(v)}$ as
	$\mathbf{\dot{z}}_{i}^{(v)} := \mathrm{MLP}_{\boldsymbol{\Theta}}^{(2)}(\mathbf{\bar{\bar{o}}}_{i}^{(v)}) \quad \text{and } \quad \mathbf{\mathring{z}}_{\text{NN}_i}^{(v)} := \verb*|sg|(\mathrm{MLP}_{\boldsymbol{\Psi}}^{(2)}(\mathbf{z}_{\text{NN}_i}^{(v)})) \,$,  respectively. Finally, same as we evaluated (\ref{eqn:normalizLIDOnline}) and (\ref{eqn:normalizLIDTarget}) for the macroscopic learning part, we also evaluate $\mathbf{\widetilde{z}}_{i}^{(v)}$ for the online and $\mathbf{\bar{z}}_{\text{NN}_i}^{(v)}$ for the target neural networks of the microscopic learning part.
	The loss function for learning the micro-fading channel characteristics is calculated over all the subcarrier reprentations $N^{\prime}$ for each view, and is written as 
	\begin{equation}
		\mathcal{L}_s:=- \frac{1}{VN'}\sum_{v=1}^{V}\sum_{i=1}^{N'} \bar{\mathbf{z}}_{\text{NN}_{i}}^{(v)} \log \left(\tilde{\mathbf{z}}_{i}^{(v)}\right)\, .\label{eqn:loss_localView} 
	\end{equation}
	
	Finally, the total loss function is computed as ${\mathcal{L}_{\text{SSL}} := \mathcal{L}_{c} + \beta \mathcal{L}_{s} \label{eqn:totalLoss}}$,
	where $\beta \in [0,1]$ controls the relevance of learning macroscopic fading versus microscopic features.
	
	\subsection{Optimization}\label{subsec:Optimization}
	In general, given the same input channel realization, multi-branch network architectures (e.g., based on Siamese as in \cite{salihu2020low}) may suffer mode collapse, resulting in both the target and online networks outputting the same constant, hence, demanding careful negative pair mining. In our case, we avoid such minima while forsaking the need for contrastive loss. We rely on expontial moving average (EMA) to build and update the target network parameters as in \cite{grill2020bootstrap}. Thus, the gradients do not backpropagate through the target network with $\boldsymbol{\Psi}$, as shown by the $\verb*|stop-gradient|$ ($\verb*|sg|$) operator in Fig. \ref{fig:SWiTArchitectureAll}. Instead, given a target decay rate $\kappa \in \left[0,1\right]$, the set of parameters $\boldsymbol{\Psi}$ is only updated via the online network after each training step as 
	$\boldsymbol{\Psi} \leftarrow \kappa \boldsymbol{\Psi}+(1-\kappa) \boldsymbol{\Theta}\,.$
	Finally, to further enhance the learning, minimize the dependency on batch size, as well as batch normalization, we perform \textit{sharpening} and \textit{centering} of the output features distribution, and freeze the target network over the first epoch, as it is suggested in \cite{9709990}. At the end of training, we only utilize the target encoder, $f_{\boldsymbol{\Psi^{\star}}}(\cdot)$, with learned parameters $\boldsymbol{\Psi^{\star}}$. Training details, and the evaluation procedure are detailed in Sec. \ref{sec:trainingDetails}.
	
	\section{Experiments and Results}\label{sec:experimentsAndResults}
	In this section, we first elaborate on the evaluation procedure and datasets. Afterwards, we assess and compare the performance of the proposed approach w.r.t. various factors. Furthermore, we provide an ablation study to build a better understanding on the usefulness of our approach.
	
	\subsection{Datasets}\label{sec:datasets}
	We evaluate the performance of self-supervised representations across various wireless propagation conditions and scenarios represented by three diverse datasets: indoors, outdoors, co-located massive MIMO, and distributed antenna systems. Next, we briefly describe these datasets.
	\subsubsection{KUL Dataset} We use \textit{ultra dense indoor MaMIMO} data pool from \cite{de2020csi} consisting of multiple datasets with varying antenna configurations and propagation characteristics, obtained in a laboratory environment. Specifically, we use three datasets: NLOS propagation and the BS with uniform rectangular array configuration (KUL-NLOS-URA-Lab), LOS and uniform linear array (KUL-LOS-ULA-Lab), and yet again, LOS environment but a distributed antenna system setup (KUL-LOS-DIS-Lab). In all cases, $N_r = 64$, $N_c^{\prime} = 100$, $f_c = 2.61 \mathrm{ GHz}$, and $R = 250\,000$.
	\subsubsection{S and HB Dataset} The S-200 and HB-200 datasets \cite{9833994} were obtained via ray-tracing and model time-varying propagation in two railway scenarios with $T=200$ realizations. The S-scenario has $M=1$ and $R=69,212$, while the HB-200 scenario has $M=8$ nodes and $R=81,200$. Both datasets have $N_{r} = 64$, $f_{c} = 3.5 \mathrm{GHz}$, $L = 4$, $W = 20 \mathrm{MHz}$, and $N_c' = 32$.
	\subsubsection{WILD Dataset} We use the WILD-v2 dataset \cite{ayyalasomayajula2020deep}, obtained indoors and suitable to understand environment adaptation capabilities. This dataset consists of CSI samples from two \textit{similar-looking} environments, Env-1 and Env-2, with $40\text{m} \times 20\text{m}$ size, and $R = 28,000$ and $R = 5,000$ samples each. Each environment has $M=6$, $f_c = 5.21 \mathrm{ GHz}$, $N_c = 234$, and $N_r = 24$. During testing, $4,000$ samples from Env-1 and $1,000$ samples from Env-2 are used, forming a test set $R_{\text{test}} = 5,000$. However, we only use Env-1 for pre-training and fine-tuning.
	
	\subsection{Self-supervised Training Details}\label{sec:trainingDetails} If not mentioned otherwise, we train SWiT without labels for $500$ epochs with $B = 512$ separately for each of the datasets. For both online and target models, $f_{\boldsymbol{\Phi}}(\cdot)$, $f_{\boldsymbol{\Psi}}(\cdot)$, $\nu_{\boldsymbol{\Phi}}(\cdot)$, and $\nu_{\boldsymbol{\Psi}}(\cdot)$, overall, we adopt architecture details of the WiT \cite{9833994}, excluding the $\mathrm{MLP}$ head, and use a base setup of one transformer block $H_{\text{blck}} = 1$ with a single attention head $H_{\text{attn}} = 1$. Since \cite{dosovitskiy2020image} has become a norm among modern transformer based models across the fields, we comply to such standard, and among other details, we use $D = 384$ instead of $D = 650$, which we commonly used in the past for WiT. The LayerNorm parameters are $\zeta=1$ and $\iota=0.0001$. We set $N_s=8$, $\widetilde{N}_c=36$, $\bar{N}_c=16$, $K_n=6$, and $K_k=3$. Due to high time complexity, we use only a small sample size (i.e., roughly 10\% of a KUL dataset) during training, which takes about 2 days on a single RTX-3070 GPU.
	
	As a solver, we use AdamW \cite{loshchilov2017decoupled} with a cosine schedule for the learning rate, $\varpi$, without restarts and a linear warm-up for $10$ epochs, and fix $\beta=0.1$. The base learning rate is $\varpi_{\text{base}} = \num{1.5e-4}$ and updates as $\varpi \triangleq \varpi_{\text{base}} B / 256$. The target network updates use $\kappa_{\text{base}} = 0.996$ updated with $\kappa \triangleq 1-\left(1-\kappa_{\text {base }}\right)(\cos (\pi u / U)+1) / 2$. We use weight decay scaled from $0.04$ to $0.4$, clip gradients at $3.0$, set $\chi_{\boldsymbol{\Psi}}=0.04$, $\chi_{\boldsymbol{\Theta}}=0.1$, and $\Lambda = 0.9$. During the first epoch, the target network is frozen. MLPs for local and global projectors consist of four layers, with the first two linear layers followed by $\mathrm{GeLU}$ non-linearity and $1,024$ nodes. The third layer is a linear bottleneck with $256$ neurons, and the last layer (i.e., the linear expander) has $25,000$ neurons for $\mathrm{MLP}^{(1)}(\cdot)$ and $1,024$ neurons for $\mathrm{MLP}^{(2)}(\cdot)$. Finally, the selected augmentations with assigned probabilities, $\mathbb{P}(Q_a)$, for $\mathcal{T}_{1}(\cdot)$, $\mathcal{T}_{2}(\cdot)$, and $\mathcal{T}_{3}(\cdot)$ are given in the Table \ref{table:Transformations_Qa_T_i}.
	\begin{table}[h]
		\caption{Transformation function $\mathcal{T}_{i}(\cdot)$ and $\mathbb{P}(Q_a)$.}
		\centering
		\scalebox{0.90}[0.90]{%
			\begin{tabular}{lccc}
				\Xhline{3\arrayrulewidth}
				\noalign{\vskip\doublerulesep
					\vskip-\arrayrulewidth}
				\Xhline{1\arrayrulewidth} 
				\multicolumn{1}{l}{$Q_a\,, \mathbb{P}(Q_a)$} 
				& \multicolumn{1}{c}{$\mathcal{T}_{1}(\cdot)$}
				& \multicolumn{1}{c}{$\mathcal{T}_{2}(\cdot)$}
				& \multicolumn{1}{c}{$\mathcal{T}_{3}(\cdot)$}                                                    
				\\[0.05cm] 
				\hline
				\noalign{\vskip\doublerulesep
					\vskip-\arrayrulewidth}
				\textit{Subcarrier selection (\textit{RSS})}                                           
				&   $1.0$                                  
				& $1.0$ 
				& $1.0$             
				\\
				\textit{Subcarrier flipping (RSF)}
				&  	\multicolumn{1}{c}{$0.4$}                                           			
				&    $0.4$                                  
				&  $0.4$               \\
				\textit{Gain offset (RGO)} 
				& \multicolumn{1}{c}{$0.2$}                                              			
				&   $0.8$                         &    $0.0$                                  \\
				\textit{Fading component (RFC)} 
				&    \multicolumn{1}{c}{$0.0$}                                        			
				&    $0.1$                   
				&       $0.0$               \\
				\textit{Sign change (RSC)} 
				&    \multicolumn{1}{c}{$0.0$}                                        			
				&    ${0.2}$                   
				&       ${0.0}$               \\
				\textit{Normalization } 
				&    \multicolumn{1}{c}{$1.0$}                                        			
				&    ${1.0}$                   
				&       ${1.0}$               \\	
				\textit{Gaussian noise } 
				&    \multicolumn{1}{c}{$0.2$}                                        			
				&    ${0.2}$                   
				&       ${0.0}$               \\
				\Xhline{3\arrayrulewidth} 
		\end{tabular}}
		\label{table:Transformations_Qa_T_i} 
	\end{table}
	\vspace{-0.40cm}
	\subsection{Evaluation Approach}\label{section:EvaluationApproach}
	We assess the performance of learned representations following the standard evaluation protocol in machine learning for self-supervised learning. More precisely, we perform linear and fine-tuning assessments for the trained models on several localization tasks, and not only that.
	\subsubsection{Linear Evaluation} For linear evaluations, we train a linear regressor, $g_{\boldsymbol{\Phi}}:\mathbb{R}^{D} \mapsto \mathbb{R}^{D_{\text{out}}}$, (or a classifier) on top of the \textit{frozen} features from $f_{\boldsymbol{\Psi}^{\star}}(\cdot)$. In the case of the regressor and the localization task, $D_{\text{out}} = 2$, i.e., location coordinates. For a just evaluation, we apply no augmentations during the fine-tuning, and if not mentioned otherwise, we report the accuracy on $N_c' = 32$ subcarriers. We commonly apply normalization, although learned embeddings are less sensitive to first-order statistics. For linear regressor, we use a one-layer MLP for $500$ epochs with $B = 128$ and AdamW with a fixed $\varpi = 3\times10^{-4}$. We also assess the quality of embeddings with a non-parametric method, i.e., the weighted nearest neighbor classifier (k-NN). Similarly, we \textit{freeze} the pre-trained model $f_{\boldsymbol{\Psi}^{\star}}(\cdot)$, and then compute and store the channel representations. Then, we apply k-NN to match the features of the input channel, and report $\verb*|top-1|$, and when relevant, $\verb*|top-5|$ accuracy. The $\verb*|top-1|$ accuracy is calculated as the proportion of correct predictions over total predictions and $\verb*|top-5|$ accuracy allows the model to output the five most likely predictions.
	\subsubsection{Fine-Tuning Evaluation} We evaluate the performance of the embeddings by fine-tuning with labeled datasets. More specifically, we add a randomly initialized MLP head with a hidden linear layer on top of the encoder, $g_{\boldsymbol{\Phi}}:\mathbb{R}^{D} \mapsto \mathbb{R}^{D_{\text{out}}}$, initialize the backbone (i.e., target encoder) with our pre-trained weights, and train the new network, $g_{\boldsymbol{\Phi}} \circ f_{\boldsymbol{\Psi}^{\star}} (\mathbf{H})$, for $\approx 150$ epochs. We use $B = 512$, $\varpi=\num{3e-4}$, and AdamW with default parameter values. Like in linear evaluation setup, the accuracy is reported on $32$ subcarriers after normalization. As other localization techniques commonly use the same batch size during the testing as during the training, it is worth noting that during the testing, we use $B=1$, as our method is batch normalization-free.
	
	While we use no labels for self-supervised training, during the fine-tuning and supervised evaluation we do stratified sampling without replacement. Furthermore, the coordinate location values, i.e., $u_{r,i}$, are scaled within $[0,1]$. The estimates are scaled back to evaluate the performance. Performance is reported in terms of mean absolute error, $\mathrm{MAE}$, and the $95-$th percentile, 
	\begin{equation}
		\begin{aligned}
			\mathrm{MAE} 
			&= \frac{ \sum_{r'=1}^{R_{\mathrm{test}}} \|\widehat{\mathbf{u}}_{r'} - \mathbf{u}_{r'} \|}{R_{\text{test}}}\;.
		\end{aligned} 
	\end{equation}
	To compare results with the reported performance from other works, we use the averaged $\mathrm{RMSE}$.	
	\subsubsection{Transfer Learning}
	To understand whether we learn generic and hence useful macroscopic along with microscopic channel characteristics, we assess the transferability across multiple datasets, environments, propagation conditions, and other wireless tasks. To do so, we use the pre-trained SWiT on the KUL-NLOS-Lab dataset and apply linear evaluation and fine-tuning on other datasets. Moreover, we apply learned representations to a relatively easy task, i.e., path-loss prediction, and compare the performance to transfer learning of a fully-supervised model trained for wireless localization. Notwithstanding our goal to solve localization, we use this example to emphasize that our method can work across multiple wireless communication prediction tasks. 
	
	\subsection{Main Results}
	\begin{figure*}[!t]%
		\centering 
		\begin{subfigure}{.32\columnwidth}
			\includegraphics[width=1.0\linewidth]{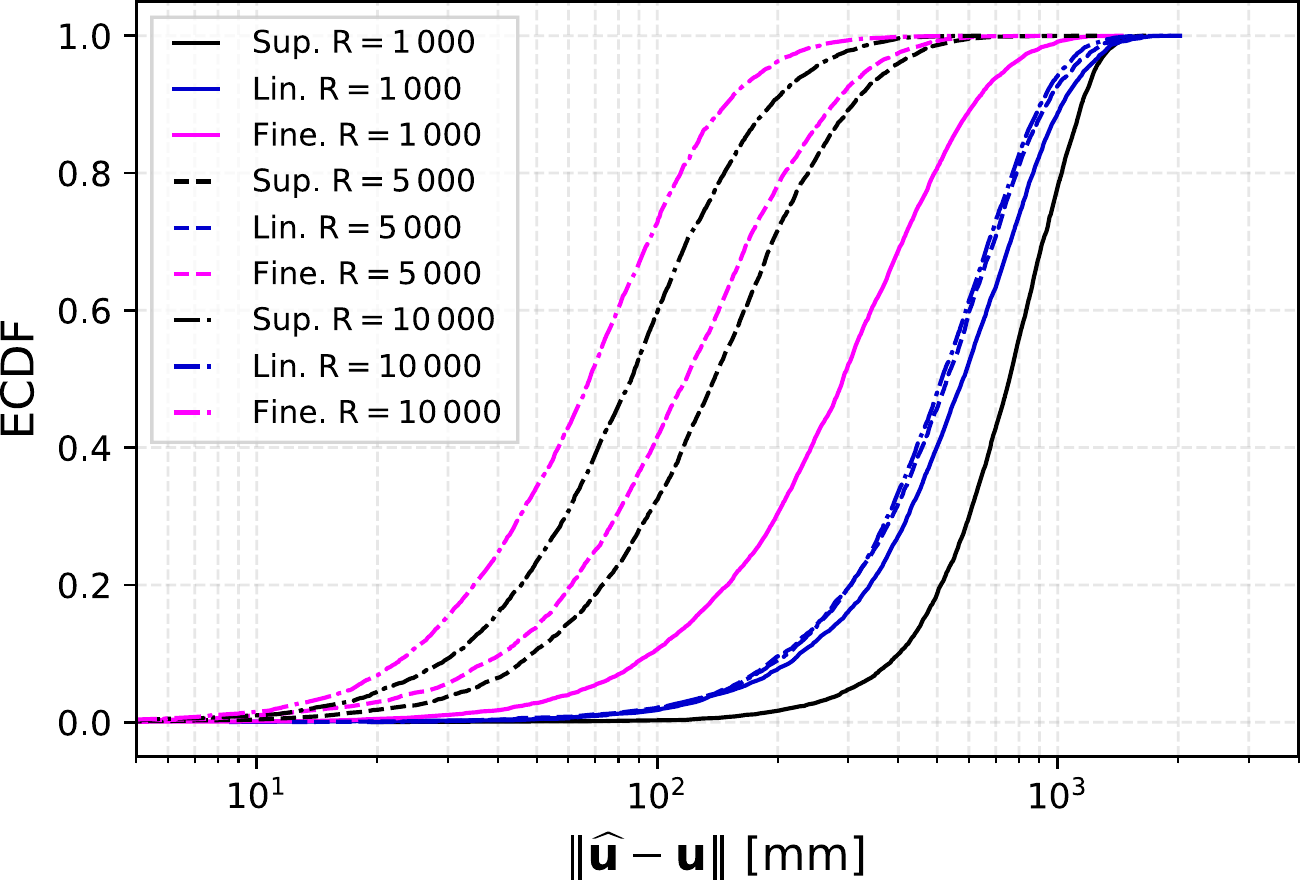}%
			\caption{KUL-NLOS-URA-Lab.}%
			\label{fig:ecdf_kul_nlos_ura_lab}%
		\end{subfigure}
		\hfill%
		\begin{subfigure}{.32\columnwidth}
			\includegraphics[width=\columnwidth]{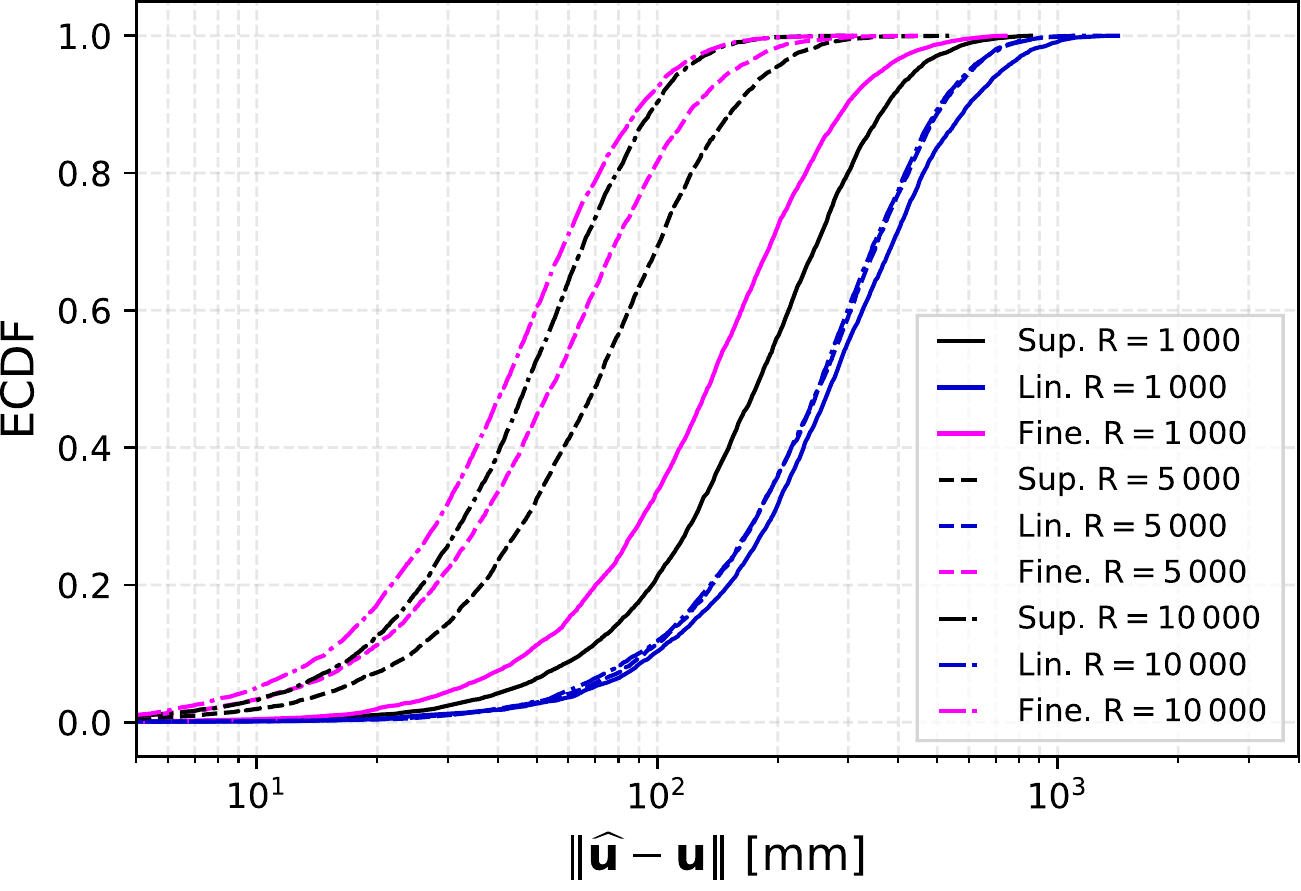}%
			\caption{KUL-LOS-ULA-Lab.}%
			\label{fig:ecdf_kul_los_dis_lab}%
		\end{subfigure}
		\hfill%
		\begin{subfigure}{.32\columnwidth}
			\includegraphics[width=\columnwidth]{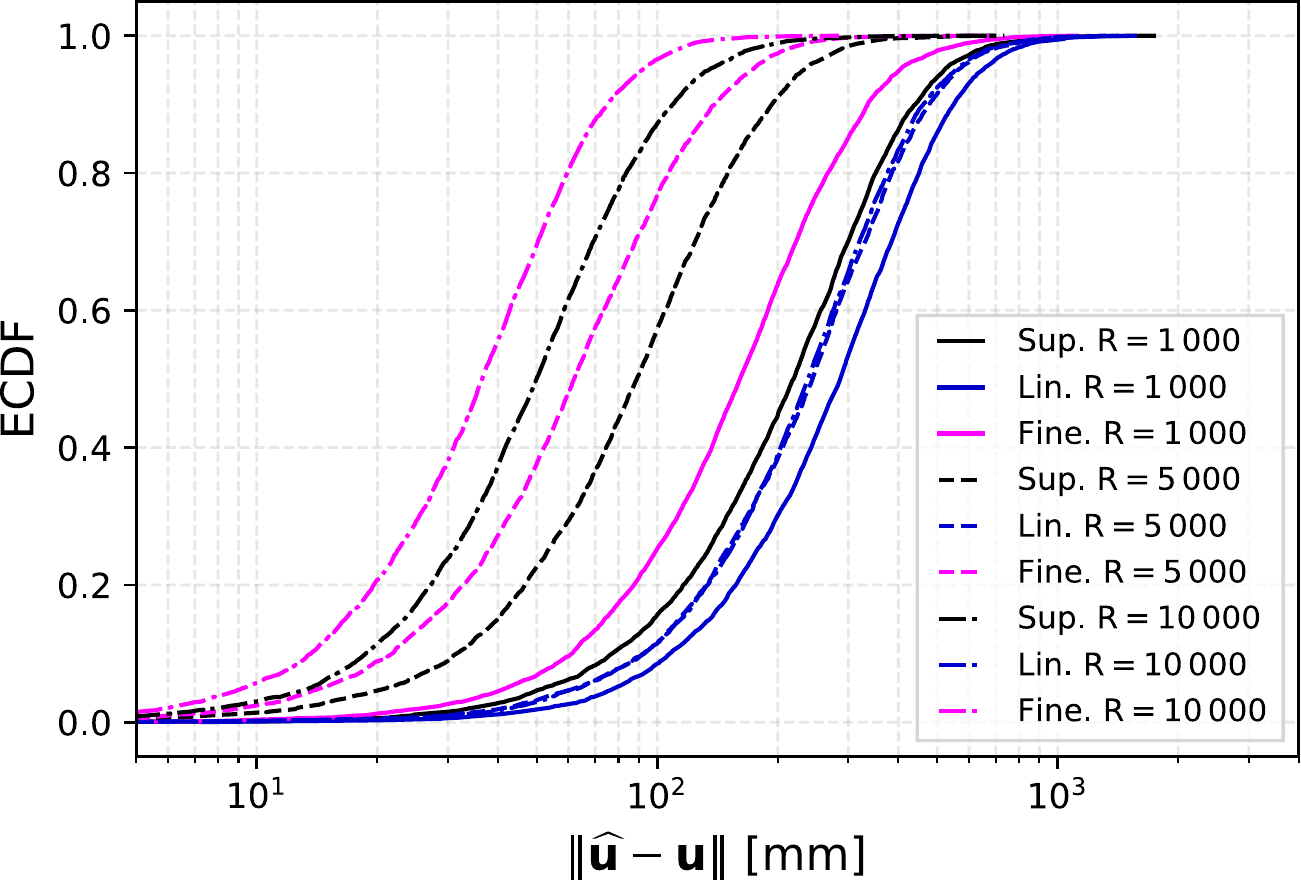}%
			\caption{KUL-LOS-DIS-Lab.}%
			\label{fig:ecdf_kul_los_ula_lab}%
		\end{subfigure}%
		\\
		\begin{subfigure}{.32\columnwidth}
			\includegraphics[width=1.0\linewidth]{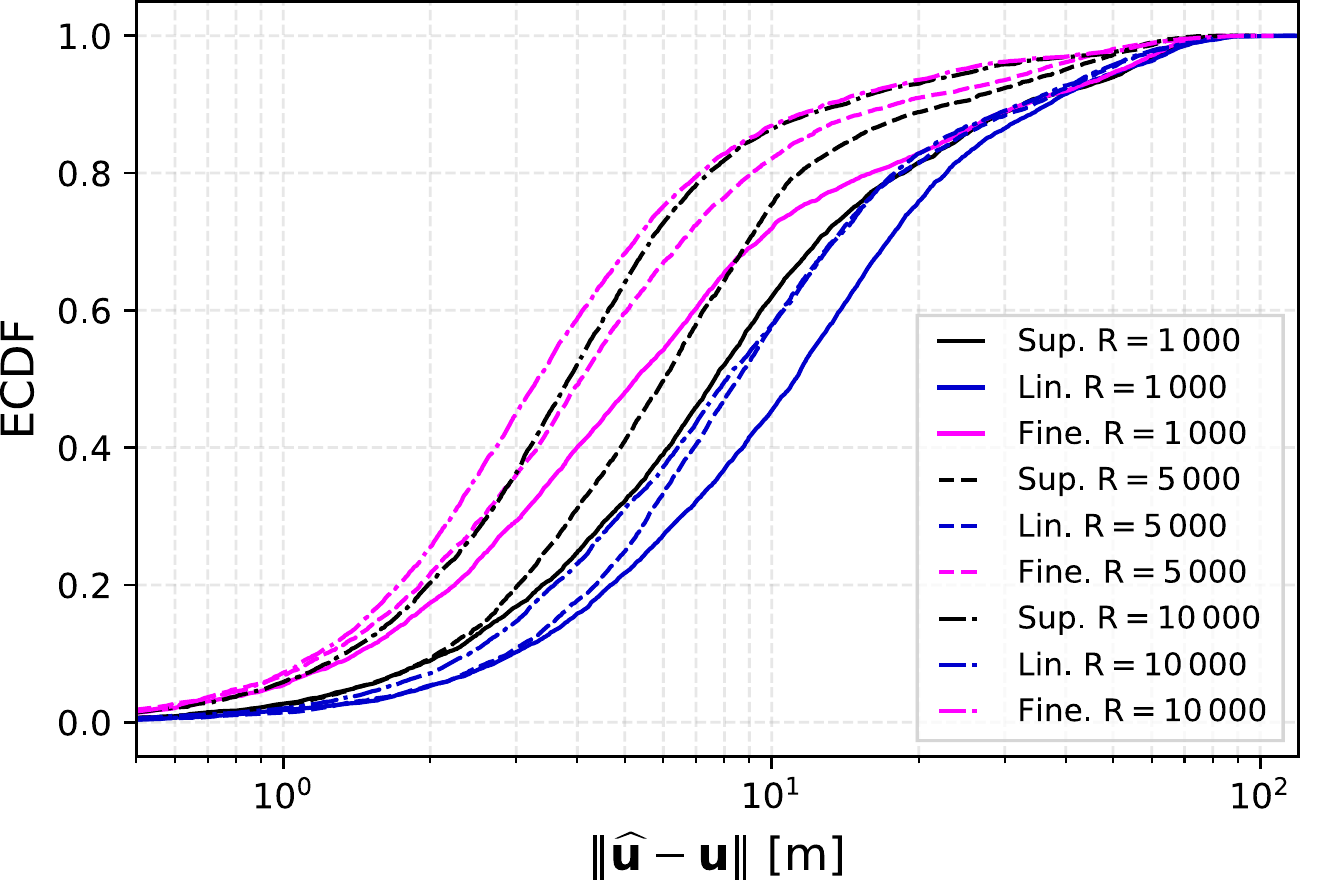}%
			\caption{S-200 co-located.}%
			\label{fig:ecdf_s_200_colocated}%
		\end{subfigure}
		\hfill%
		\begin{subfigure}{.32\columnwidth}
			\includegraphics[width=\columnwidth]{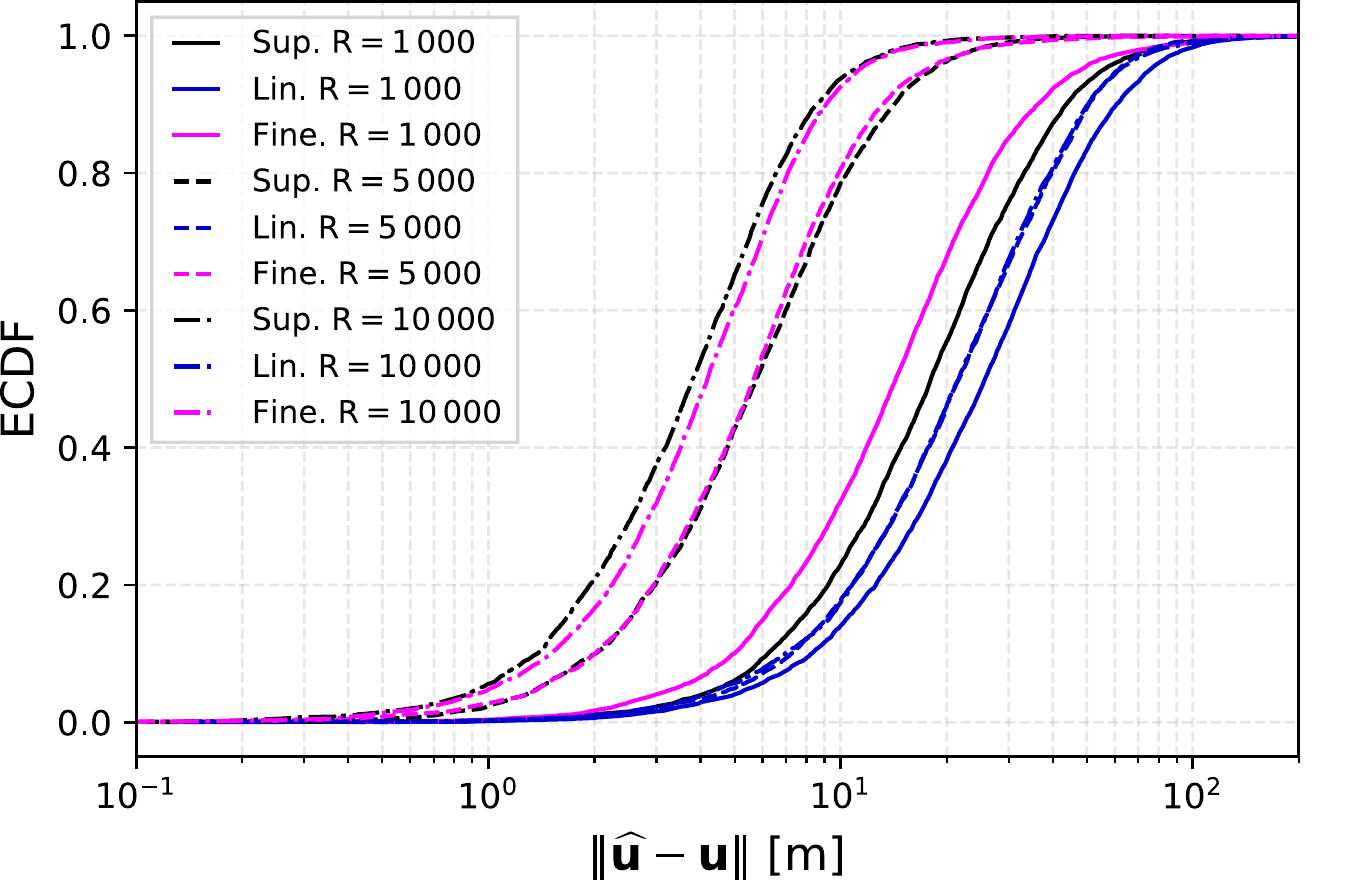}%
			\caption{HB-200 DIS.}%
			\label{fig:ecdf_hb_200_dis}%
		\end{subfigure}
		\hfill%
		\begin{subfigure}{.32\columnwidth}
			\includegraphics[width=\columnwidth]{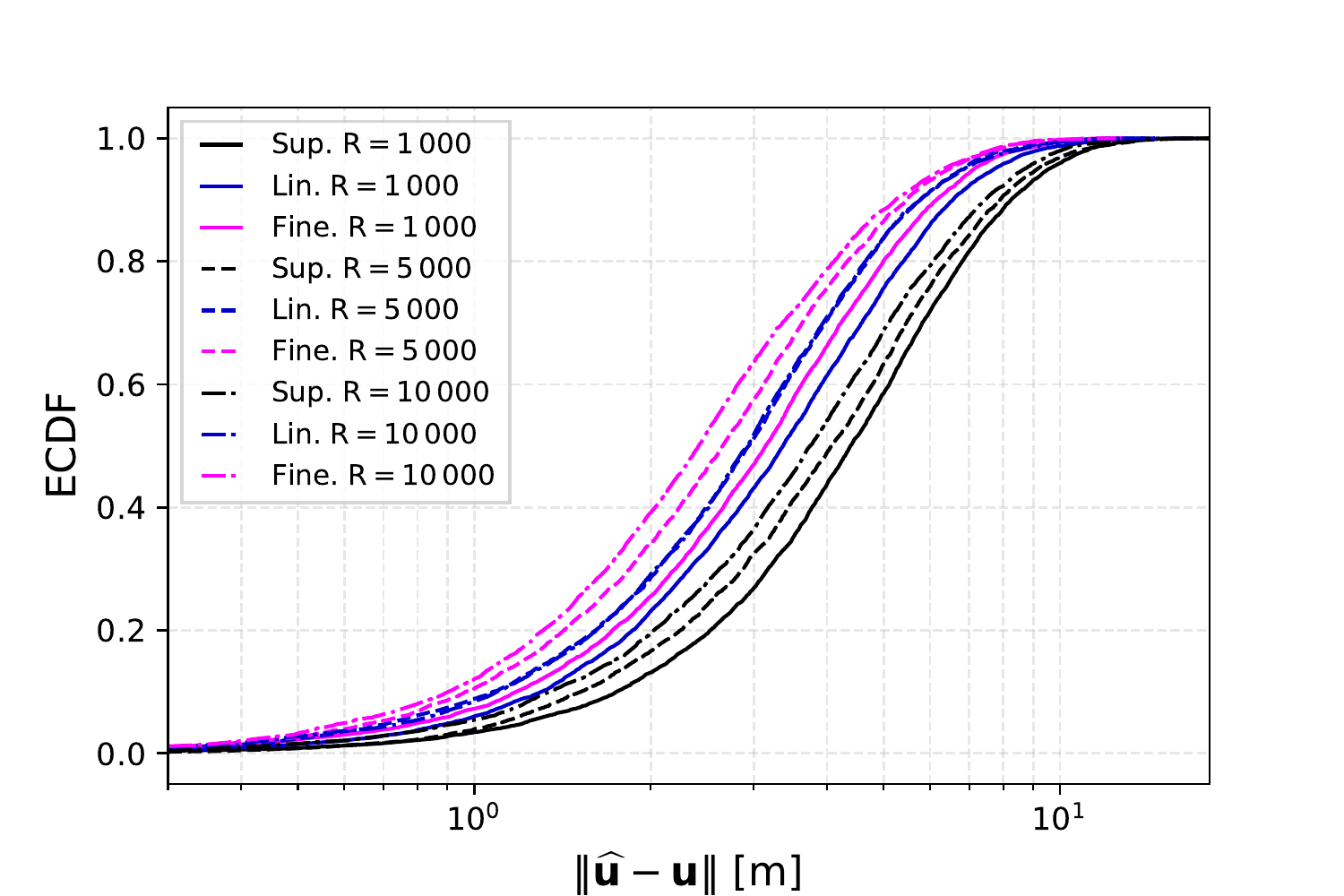}%
			\caption{WILD DIS.}%
			\label{fig:ecdf_WILD_dis}%
		\end{subfigure} 
		\caption{Self-supervised and fine-tuning outperform fully-supervised approach. a) NLOS, b) LOS-ULA, c) LOS-DIS, d) Outdoors S, d) Outdoors HB and DIS, and e) Indoors and DIS (WILD-v2). Fully-supervised and fine-tuned models trained for $150$ epochs. Linear evaluation for $500$ epochs.}
		\label{fig:ecdf_error_kul_scenarios_swit}
	\end{figure*}
	
	\subsubsection{Localization Accuracy}
	In Fig. \ref{fig:ecdf_error_kul_scenarios_swit}, we show the performance while varying the number of training samples for the datasets described in Sec. \ref{sec:datasets}. As mentioned earlier, we assess the performance under three setups: when using a linear regressor, fine-tuning the pre-trained SWiT encoder, or training a randomly initialized encoder. For the data regimes, we use $R = \{1\,000, 5\,000, 10\,000\}$ for training and $R_{\text{test}} = 5\,000$ without replacement for testing. As can be seen from the figure, the pre-trained models outperform the fully-supervised case in \textit{all} three data regimes. We observe that when a very small amount of training samples is available, in certain conditions, we can outperform the fully-supervised training even with a linear regressor. This behavior is especially observed in Fig. \ref{fig:ecdf_WILD_dis}, i.e., for the WILD-v2 dataset. To illustrate the impact of \textit{post-normalization}, in Fig. \ref{fig:ecdf_kul_nlos_ura_lab} we show the linear regressor trained on self-supervised embeddings is more robust than fully-supervised without normalization when $R = 1,000$.
	
	\subsubsection{Spot-Localization}
	We investigate unsupervised spot-localization, i.e., the ability of the model to \textit{cluster} correlated channel representations while preserving the spatial neighborhood. To do so, we consider KUL, S-200, and HB-200 datasets. Since datasets in KUL are obtained from a scenario that is divided into four sub-regions, we consider $\widehat{C}_{\text{spot}} = 4$ spots. On the other hand, for S-200 and HB-200 datasets, we have $\widehat{C}_{\text{spot}} = 360$ and $\widehat{C}_{\text{spot}} = 406$, respectively. In Table \ref{table: results_table_classification_KUL_and_SHB}, we can read that embeddings from SWiT encoder can achieve near perfect accuracy for KUL datasets. This evaluation procedure clearly indicates that the pre-trained model yields remarkably better representations compared to embeddings of the a randomly initialized WiT.
	\begin{table}[!h]
		\caption{Random weights versus SWiT.}
		\centering
		\scalebox{1.00}[1.00]{%
			\begin{tabular}{lccccccccc}
				\Xhline{3\arrayrulewidth}
				\noalign{\vskip\doublerulesep
					\vskip-\arrayrulewidth}
				\Xhline{1\arrayrulewidth} 
				& \multicolumn{1}{c}{ KUL-NLOS } & & \multicolumn{1}{c}{ KUL-LOS } & & \multicolumn{1}{c}{ KUL-LOS-DIS } & \multicolumn{2}{c}{ S-200 } & \multicolumn{2}{c}{ HB-200 } \\
				\cline {2-10}
				\textbf{Method} & $\, \uparrow$ Top-1 & & $\, \uparrow$ Top-1 & & $\, \uparrow$ Top-1  & $\, \uparrow$ Top-1 & $\, \uparrow$ Top-5 & $\, \uparrow$ Top-1 & $\, \uparrow$ Top-5 \\
				\hline
				Random & $23.7$ & & $3.59$  & & $4.13$  & $0.296$ & $1.545$ & $0.27$ & $1.25$ \\
				\rowcolor{LightCyan} SWiT & $99.99$ & & $99.99$  && $99.99$ & $18.70$ & $52.38$ & $37.38$ & $82.85$ \\
				\\[-0.51cm]
				\Xhline{3\arrayrulewidth} 
		\end{tabular}}
		\label{table: results_table_classification_KUL_and_SHB} 
	\end{table}
	
\begin{figure}[!h] 
		\centering 
		\subfloat[NLOS Random. \label{fig:nlos_ura_lab_rand}]{%
			\includegraphics[width=0.20\linewidth]{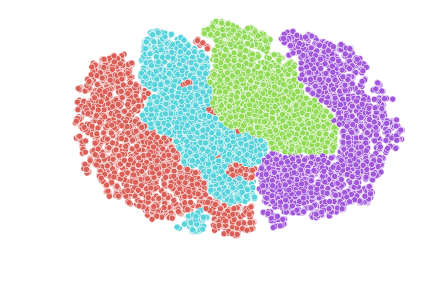}}
		\hfill
		\subfloat[NLOS SWiT. \label{fig:nlos_ura_lab_swit}]{%
			\includegraphics[width=0.20\linewidth]{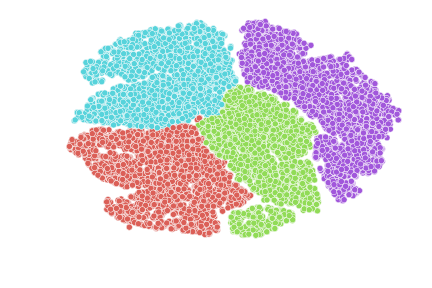}}
		\hfill
		\subfloat[LOS-DIS Random. \label{fig:los_dis_lab_rand}]{%
			\includegraphics[width=0.20\linewidth]{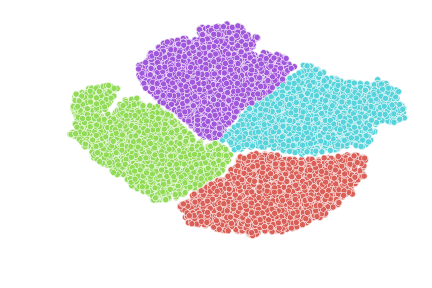}}
		\\
		\subfloat[LOS-DIS SWiT. \label{fig:los_dis_lab_swit}]{%
			\includegraphics[width=0.20\linewidth]{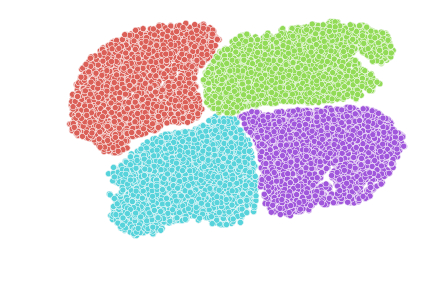}}	
		\hfill
		\subfloat[LOS Random. \label{fig:los_ula_lab_rand}]{%
			\includegraphics[width=0.20\linewidth]{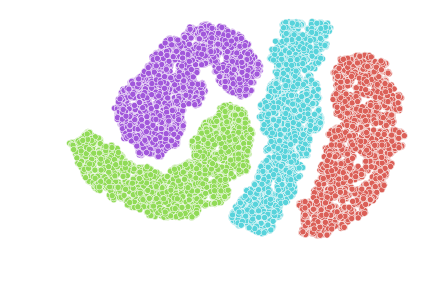}}
		\hfill
		\subfloat[LOS SWiT. \label{fig:los_ula_lab_swit}]{%
			\includegraphics[width=0.20\linewidth]{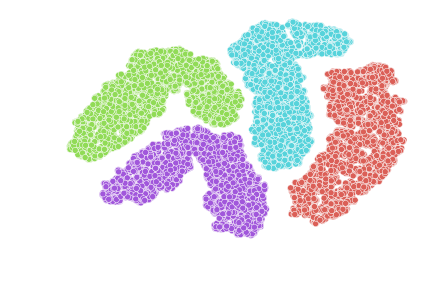}}  
		\caption{t-SNE embeddings of the channel with randomly initialized weights compared to SWiT. a) and b) depicts the case of NLOS, c) and d) LOS-DIS, e) and f) LOS-ULA case.}
		\label{fig:tSne_embeddings_KUL}
	\end{figure}

	At last, in Fig. \ref{fig:tSne_embeddings_KUL}, in the interest of visualization, we depict the computed two-dimensional t-SNE \cite{van2008visualizing} embeddings of the KUL dataset. We can easily observe that for the LOS propagation conditions, randomly initialized WiT and pre-trained models can yield equally qualitative representations. However, when in NLOS, SWiT significantly improves the quality of representations. This is shown in Fig. \ref{fig:nlos_ura_lab_rand} and Fig. \ref{fig:nlos_ura_lab_swit}, where each color represents a class, i.e., a \textit{spot}.
	
	\subsection{Other Works}
	There has been a great amount of DNN-based localization techniques proposed over the years. Regardless, no such approach operates on self-supervised embeddings. Nonetheless, we compare the localization performance to the reported results of the selected techniques by \cite{10018917}. While we observe some ambiguity in evaluation metrics and dataset selection procedure, we attempt to comply and report the error for the NLOS case in Table \ref{table: results_comparision_other_works}. From the table, we can observe that pre-trained WiT alongside randomly initialized WiT has a remarkable performance compared to other works, even when trained only for $150$ epochs, i.e., almost seven times less computing time. Furthermore, we outperform some localization techniques in small data regimes whilst training only a linear regressor on \textit{frozen} parameters. In all cases $R_{\text{test}}=5\,000$. 
	\begin{table}[h]
		\caption{Localization error in $\left[\text{mm}\right]$.}
		\centering
		\scalebox{0.90}[0.90]{%
			\begin{tabular}{lccccc}
				\Xhline{3\arrayrulewidth}
				\noalign{\vskip\doublerulesep
					\vskip-\arrayrulewidth}
				\Xhline{1\arrayrulewidth} 
				& \multicolumn{1}{c}{$R=1\,000$} & & \multicolumn{1}{c}{$R=5\,000$}  & & \multicolumn{1}{c}{$R=10\,000$} \\
				\textbf{Method} & $\, \downarrow \mathrm{MSE}$ & & $\, \downarrow \mathrm{MSE}$ & & $\, \downarrow \mathrm{MSE}$ \\
				\hline Sun19 \cite{sun2019fingerprint} & $866.14$ & & $533.94$ & & $392.13$\\
				Arnold19 \cite{arnold2019novel}  & $785.42$ & & $322.67$ & & $204.54$ \\
				Chin20 \cite{9136584} &$660.77$ & & $248.25$ & & $118.75$ \\
				Bast20 \cite{de2020csi} & $826.42$ & & $483.28$ & & $356.61$ \\
				MHSA22 \cite{liu2022mhsa} & $702.96$ & & $351.104$ & & $243.07$ \\
				Wang18 \cite{8468057} & $1098.27$ & & $728.89$ & & $626.95$ \\
				Hoang20 \cite{hoang2020cnn} & $820.75$ & & $463.94$ & & $342.99$ \\
				AAResCNN \cite{10018917} & $598.96$ & & $193.65$ & & $108.34$ \\
				\rowcolor{LightCyan} WiT \cite{9833994} & $427.057$ & & $190.588$ & & $112.502$ \\
				\\[-0.50cm]
				\hline 
				\\[-0.45cm]
				\shortstack{\underline{Ours:} \\ SWiT $+$ Linear} & $570.42$ & & $467.6$ & & $459.08$ \\
				\\[-0.55cm]
				\rowcolor{LightCyan}\shortstack{SWiT $+$ Fine-tuning}   & $366.90$ & & $156.96$ & & $88.45$ 
				\\
				\\[-0.50cm]
				\Xhline{3\arrayrulewidth}
		\end{tabular}}
		\label{table: results_comparision_other_works} 
	\end{table}
	\subsection{Transfer Learning to Other Datasets}
	We evaluate the transferability of the pre-trained model to other datasets in Table \ref{table:results_transferability_all}, when $R = R_{\text{test}} = 5\,000$ samples. We can observe that despite noticeable, and sometimes even significant, variations in localization performance, overall, one can pre-train a model only on a single dataset while achieving relatively good localization performance across all environments and antenna configuration setups. For instance, a model trained only in the KUL-NLOS-URA dataset can achieve the same spot estimation performance on other KUL datasets and yields less than $2\%$ degradation in $\verb|top-1|$ accuracy for S and HB datasets. Additionally, models trained on ray-tracing channels (synthetic data) can sufficiently transfer learning to realistic environments. For example, the models trained on S and HB environments can achieve similar linear location estimates when evaluated on the KUL-LOS-DIS dataset. 
	
	\begin{table*}[h]
		\caption{Transferability across different datasets for \textit{spot} estimation and location mapping.}
		\centering
		\scalebox{0.90}[0.90]{%
			\begin{tabular}{lcccccccccc}
				\Xhline{3\arrayrulewidth}
				\noalign{\vskip\doublerulesep
					\vskip-\arrayrulewidth}
				\Xhline{1\arrayrulewidth} 
				& \multicolumn{2}{c}{KUL-NLOS-URA}                                            &
				\multicolumn{2}{c}{KUL-LOS-ULA}                                             & \multicolumn{2}{c}{KUL-LOS-DIS}                                             & \multicolumn{2}{c}{S-200}                                                   & \multicolumn{2}{c}{HB-200}                                                  
				\\ \hline
				\noalign{\vskip\doublerulesep
					\vskip-\arrayrulewidth}
				\hline
				\multicolumn{11}{l}{\vspace{-0.1cm}\textit{\textbf{Classifier:}}}                   \\[-0.02cm] 
				& \multicolumn{2}{c}{$\mathrm{Top-1}$} & \multicolumn{2}{c}{$\mathrm{Top-1}$} & \multicolumn{2}{c}{$\mathrm{Top-1}$} & \multicolumn{1}{c}{$\mathrm{Top-1}$} & \multicolumn{1}{c}{$\mathrm{Top-5}$} & \multicolumn{1}{c}{$\mathrm{Top-1}$} & \multicolumn{1}{c}{$\mathrm{Top-5}$} \\
				\cline{2-11} 
				KUL-NLOS-URA 
				&  \multicolumn{2}{c}{$\textbf{99.84}$}                                               
				&  \multicolumn{2}{c}{$99.98$}                                     
				&  \multicolumn{2}{c}{$99.82$}                                      
				&   $16.1$                                   &   $44.34$                 
				&   $20.68$                                   & $53.58$              \\
				KUL-LOS-ULA 
				&  	\multicolumn{2}{c}{$98.86$}                                           			
				&  \multicolumn{2}{c}{$\textbf{99.98}$}                                     
				&  \multicolumn{2}{c}{$99.43$}                                      
				&    $17.12$                                  &   $46.16$                       
				&    $17.8$                                  &  $49.18$               \\
				KUL-LOS-DIS 
				& \multicolumn{2}{c}{$99.10$}                                              			
				&  \multicolumn{2}{c}{$99.98$}                                     
				&  \multicolumn{2}{c}{$\textbf{99.82}$}                                      
				&   $13.04$                                   &  $40.14$                      
				&   $20.4$                         &    $56.14$                                  \\
				S-200       
				&  \multicolumn{2}{c}{$99.12$}                                     			
				&  \multicolumn{2}{c}{$99.96$}                                     
				&  \multicolumn{2}{c}{$99.52$}           
				&    $\textbf{17.69}$           &    $\textbf{47.64}$                         
				&     $12.1$                                 &  $35.74$                     \\
				HB-200      
				&    \multicolumn{2}{c}{$99.48$}                                        			
				&  \multicolumn{2}{c}{$99.98$}                                     
				&  \multicolumn{2}{c}{$99.52$}                                      
				&    $12.08$                                  &     $38.24$                      
				&    $\textbf{21.05}$                   &       $\textbf{56.14}$               \\ \Xhline{1\arrayrulewidth} 
				\\[-0.55cm]
				\multicolumn{11}{l}{\vspace{-0.1cm}\textit{\textbf{Linear Regression:}}}                                                                                                                                                                                                                                                                                                                                                          \\[-0.01cm] 
				& \multicolumn{1}{c}{$\mathrm{MAE}$}   & \multicolumn{1}{c}{$95-$th}          & \multicolumn{1}{c}{$\mathrm{MAE}$}   & \multicolumn{1}{c}{$95-$th}          & \multicolumn{1}{c}{$\mathrm{MAE}$}   & \multicolumn{1}{c}{$95-$th}          & \multicolumn{1}{c}{$\mathrm{MAE}$}   & \multicolumn{1}{c}{$95-$th}          & \multicolumn{1}{c}{$\mathrm{MAE}$}   & \multicolumn{1}{c}{$95-$th}          \\
				\cline{2-11} 
				KUL-NLOS-URA 
				&       $\textbf{453.894}$             & $\textbf{908.275}$                          &       ${282.673}$             		& ${611.641}$    
				&       $274.468$   					&  $567.503$                    
				&      	$15.64$                         &  $49.53$                                    &       $23.922$                        &  $56.047$                           \\
				KUL-LOS-ULA 
				&    $582.721$                         &  $1098.16$                                    &   $\textbf{321.528}$              &   $\textbf{669.007}$                     
				&   $286.267$                          &   $590.466$                                   &   $16.702$                           &    $49.18$                                  &   $27.82$                            &   $67.53$                          \\
				KUL-LOS-DIS 
				& 	$557.30$                           &  $1080.42$                                   & ${267.887}$   					&  ${573.141}$ 
				& $\textbf{281.402}$   					&  $\textbf{596.043}$  
				&  $18.068$                            &   $50.48$                                   &  $27.68$                           &  $63.269$                           \\
				S-200       
				&    $551.957$                         &     $1068.88$                                 &    $ 278.177 $                       &       $584.959$                               &  $273.491$                  			&    $565.191$          
				&  $\textbf{14.56}$                    &   $\textbf{51.67}$                       
				&   $   37.24$                     &        $78.422$                            \\
				HB-200      
				&  $549.967$                           &       $1065.52$                               &   $273.01$                           &       $582.07$                               &   $277.331$                          &      $569.638$                                &  $18.69$                             &       $50.140$                               &   $\textbf{26.25}$                &  $\textbf{63.500}$                            
				\\
				\Xhline{3\arrayrulewidth} 
		\end{tabular}}
		\label{table:results_transferability_all} 
	\end{table*}
	\subsection{Transfer Learning to Other Wireless Tasks}
	As we mentioned in Sec. \ref{section:EvaluationApproach}, we also evaluate the transferability of supervised and self-supervised models to a task different from localization, i.e., path-loss prediction. In Table \ref{table:TransferLearningPathLoss}, we show that a supervised model trained for channel-to-location mapping cannot transfer well to path-loss prediction. On the other hand, a model trained with a linear pathloss prediction head on top of the frozen features, can achieve accuracy similar to the fully-supervised model trained on the same task. We split the dataset into $0.8$ for training and $0.2$ for evaluation and testing.
	\begin{table}[!t]
		\caption{Transfer Learning for path-loss prediction in [dB].}
		\centering
		\scalebox{1.00}[1.00]{%
			\begin{tabular}{lcc}
				\Xhline{3\arrayrulewidth}
				\noalign{\vskip\doublerulesep
					\vskip-\arrayrulewidth}
				\Xhline{1\arrayrulewidth} 
				& \multicolumn{2}{c}{S-200 } \\
				\cline { 2 - 3 }
				\textbf{Method} & $\, \downarrow$ MAE & $\, \downarrow$ $95-$th \\
				\hline 
				Fully-Supervised & $5.917$ & $18.493$ \\
				Transfer-learning (Linear) & $16.08$ & $31.682$  \\
				\rowcolor{LightCyan} SWiT+Linear & $6.594$ & $18.426$  \\				
				\\[-0.51cm]
				\Xhline{3\arrayrulewidth}
		\end{tabular}}
		\label{table:TransferLearningPathLoss} 
	\end{table}
	\subsection{Impact of Channel Transformations}
	In Table \ref{table:impact_of_transforms_v2} we show the impact of channel transformations discussed in Sec. \ref{sec:channelAugmentations}. We assess the performance by removing a transformation from the set $\mathcal{Q}$. More precisely, $\mathcal{T}_{\text{aug}}^{(1)}$ corresponds to the complete set of augmentations. When we use $\mathcal{T}_{\text{aug}}^{(2)}$, it matches the subset without \textit{RSF}, $\mathcal{T}_{\text{aug}}^{(3)}$ indicates the subset without \textit{flipping} and \textit{RGO}. Further, when $\mathcal{T}_{\text{aug}}^{(4)}$, we add no \textit{fading component}, i.e., we remove \textit{RFC} too. Finally, for $\mathcal{T}_{\text{aug}}^{(5)}$, we use only \textit{RSS} and \textit{normalization}. In general, we observe that channel transformations slightly degrade the performance when assessed in KUL datasets for $R = R_{\text{test}} = 5\,000$. However, we observe improvements in the HB-200 scenario, and further gains can be emphasized when larger training dataset sizes are used during the fine-tuning. Overall, we can achieve high-quality self-supervised channel representations while unconstrained on the view selection function.
	\begin{table}[h]
		\caption{Impact of channel transforms, $\mathcal{T}_{\text{aug}}(\cdot)$.}
		\centering
		\scalebox{0.95}[0.95]{%
			\begin{tabular}{lcccccccccccc}
				\Xhline{3\arrayrulewidth}
				\noalign{\vskip\doublerulesep
					\vskip-\arrayrulewidth}
				\Xhline{1\arrayrulewidth} 
				& \multicolumn{2}{c}{KUL-NLOS-URA} & \multicolumn{2}{c}{HB-200} &&& \multicolumn{2}{c}{KUL-NLOS-URA}  && \multicolumn{2}{c}{HB-200}  &                                                   
				\\ \hline
				\noalign{\vskip\doublerulesep
					\vskip-\arrayrulewidth}
				\hline
				\multicolumn{2}{l}{\vspace{-0.1cm}\textit{\textbf{Classification:}}}   &&&& \multicolumn{2}{l}{\vspace{-0.1cm}\textit{\textbf{Regression:}}}                                                                                     \\ \\[-0.40cm]
				& \multicolumn{2}{c}{$\mathrm{Top-1}$} & \multicolumn{1}{c}{$\mathrm{Top-1}$} & \multicolumn{1}{c}{$\mathrm{Top-5}$} &&&  \multicolumn{1}{c}{$\mathrm{MAE}$}   & \multicolumn{1}{c}{$95-$th} &&  \multicolumn{1}{c}{$\mathrm{MAE}$} & \multicolumn{1}{c}{$95-$th}          \\
				\cline{2-5} 
				\cline{8-13}  
				$\mathcal{T}_{\text{aug}}^{(1)}$ 
				&  \multicolumn{2}{c}{${99.64}$}                                               
				&   $21.12$                                   & $56.16$  && 					&       ${445.2}$           & ${901.0}$                          
				&&      ${23.22}$                 & ${57.05}$         \\
				$\mathcal{T}_{\text{aug}}^{(2)}$ 
				&  	\multicolumn{2}{c}{$99.83$}                                           			
				&    $18.66$                                  &  $53.24$  && 					&    $446.537$           		&  $908.679$                                   
				&&   ${23.92}$         &${58.90}$                                 \\
				$\mathcal{T}_{\text{aug}}^{(3)}$ 
				& \multicolumn{2}{c}{$99.76$}                                              			
				&   $19.46$                         &    $52.9$      && 	
				& $435.319$                        &$864.965$                                   && ${23.60}$   					&  $58.66$            \\                      
				$\mathcal{T}_{\text{aug}}^{(4)}$    
				&    \multicolumn{2}{c}{$99.72$}                                        			
				&    ${20.18}$                   &       ${54.06}$  && 	   
				& $429.302$                    &     $857.742$                                 &&    $23.54$      				&  $57.26$           \\
				$\mathcal{T}_{\text{aug}}^{(5)}$    
				&    \multicolumn{2}{c}{$99.96$}                                        			
				&    ${20.08}$                   &       ${53.94}$   &&
				& $420.821$             &       $833.4$                                  
				&&   $24.11$             &       $57.38$ 
				\\ 
				\Xhline{3\arrayrulewidth} 
		\end{tabular}}
		\label{table:impact_of_transforms_v2} 
	\end{table}
	\subsection{Relevance of Fading Characteristics}
	In Table \ref{table:BetaImpactLoss}, we show the relevance of learning macroscopic versus microscopic channel characterisitcs. To do so, we investigate the accuracy for $\beta = \{0.0001, 0.1, 0.3, 0.5, 0.7, 0.9\}$. We use $R = R_{\text{test}} = 5\,000$, and for the linear regression case, we train the linear head for $500$ epochs. We can observe that relying on large-scale channel features is more relevant for the localization task. Performance degrades as we start to weight equally the loss for two learning tasks.
	\begin{table}[h]
		\caption{Impact of $\beta$.}.
		\centering
		\scalebox{0.95}[0.95]{%
			\begin{tabular}{lccc}
				\Xhline{3\arrayrulewidth}
				\noalign{\vskip\doublerulesep
					\vskip-\arrayrulewidth}
				\Xhline{1\arrayrulewidth} 
				& \multicolumn{3}{c}{KUL-NLOS-URA} \\
				\cline { 2 - 4 }
				$\beta$ & $\, \uparrow$ Top-1 & $\, \downarrow$ MAE & $\, \downarrow$ $95-$th \\
				\hline 
				$0.0001$ & $99.86$ & $452.287$ & $903.783$ \\				
				$0.1$ & $99.86$ & $442.602$ & $879.146$ \\
				$0.3$ & $99.88$ & $437.764$   & $879.534$ \\
				$0.5$ & $99.86$ & $455.951$ & $905.976$  \\	
				$0.7$ & $99.84$ & $452.186$ & $900.442$  \\
				$0.9$ & $99.80$ & $458.57$ & $920.516$  \\							
				\Xhline{3\arrayrulewidth}
		\end{tabular}}
		\label{table:BetaImpactLoss} 
	\end{table}	
	\subsection{Varying Input Channel Representation}\label{sub:varyInput}
	In general, the main goal of WiT is to maintain and exploit the per-subcarrier channel structure through a transformer architecture \cite{9833994}. However, as the channel can also be represented over space and frequency, it is reasonable to explore alternative \textit{slicing} strategies to feed the sequence of resulting slices (a.k.a tokens) into a transformer block. One motivation for investigating alternative \textit{slicing} strategies is to incorporate the temporal dynamics of the environment into the learning process. By accounting for multiple geo-tagged channel estimates obtained over time from the same location and capturing dependencies between different tokens, we can improve the accuracy and robustness of the model. Further, we can take advantage of a prior knowledge on antenna array configuration. For instance, dividing the array into sub-arrays commonly helps to reduce the spatial correlation between the sub-arrays, which benefits the learning of the underlying patterns between the resulting representations. Among multiple investigated designs, in this part, we show results for three most promising architecture choices which are illustrated in Fig. \ref{fig:wit_extended_attention_designs} and explained below. For the results that are depicted in Fig. \ref{fig:time_frequency_antenna_separated_atttention}, we train the models for $1800$ epochs on $R \approx 55\,000$ samples and report the normalized MSE (NMSE) in dB.
	\subsubsection{WiT-E-TF}\label{subsub:WiT-E-TF} 
\begin{figure*}[!t]%
		\centering 
		\begin{subfigure}{.325\columnwidth}
			\includegraphics[width=0.85\linewidth]{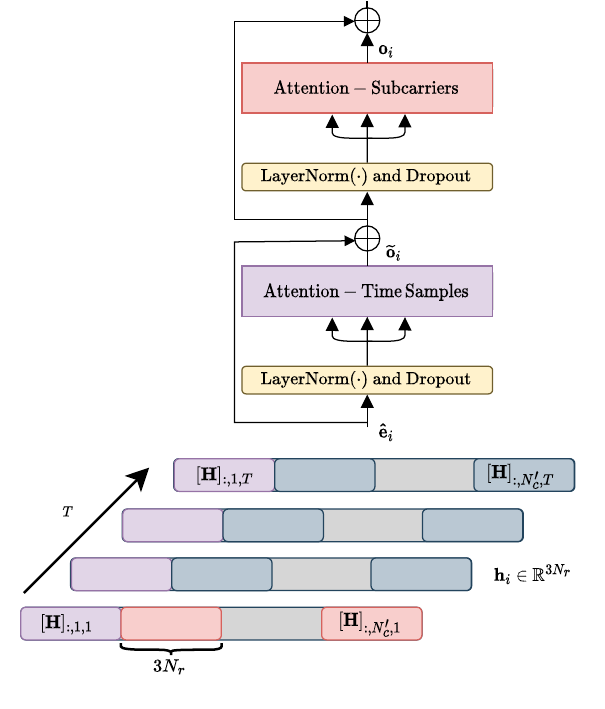}%
			\vspace{-0.25cm}\caption{Time-frequency (WiT-E-TF).}%
			\label{fig:WiTE_time_subcarrier_attention}%
		\end{subfigure}
		\hfill%
		\begin{subfigure}{.325\columnwidth}
			\includegraphics[width=0.85\columnwidth]{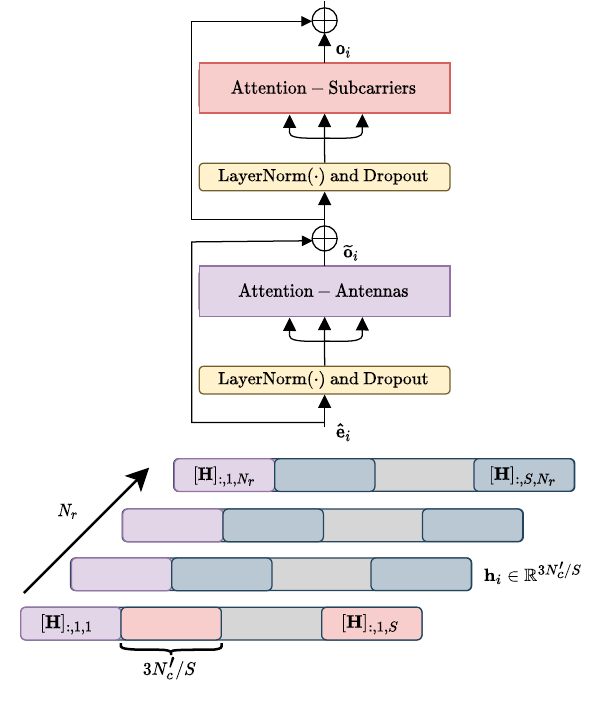}%
			\vspace{-0.25cm}\caption{Antenna-frequency (WiT-E-AF).}%
			\label{fig:WiTE_antenna_subcarrier_attention}%
		\end{subfigure}
		\hfill%
		\begin{subfigure}{.325\columnwidth}
			\includegraphics[width=0.85\columnwidth]{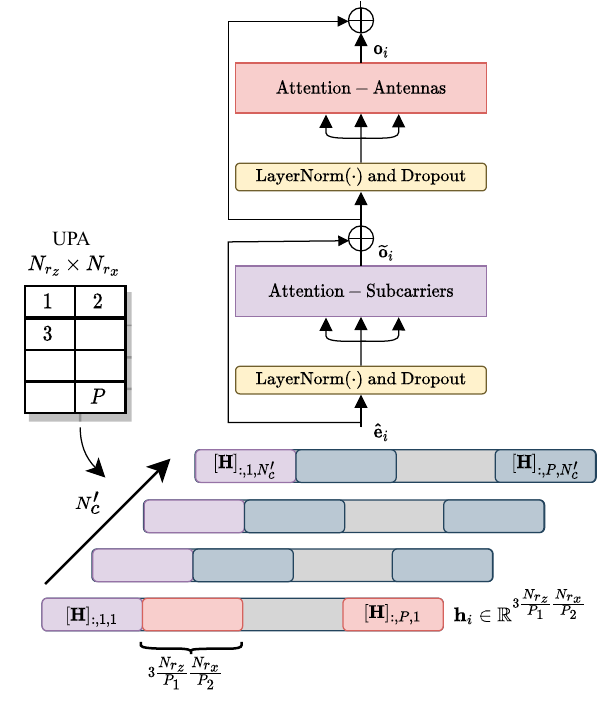}%
			\vspace{-0.25cm}\caption{Frequency-antenna (WiT-E-UPA).}%
			\label{fig:WiTE_subcarrier_antenna_upa_attention}%
		\end{subfigure}%
		\caption{Studies on the impact of input channel representation. Three approaches: a) time and subcarrier attention, b) subcarrier and antenna attention, and c) array motivated attention.}
		\label{fig:wit_extended_attention_designs}
	\end{figure*}

Here, we aim to exploit the temporal variations of the environment by stacking $T$ channel matrices obtained from the same location and viewing the channel as $\mathbf{H}_r \in \mathbb{C}^{N_r \times N_c' \times T}$. The input representation to this design is still $\mathbf{h}_{i} \in \mathbb{R}^{1\times 3N_r}$. However, considering $T$ representations would require $C = (N_c'T + 1)$ computations. To address this issue, we propose to employ a factorized dual-attention approach, resulting in a reduction of computational complexity to $(N_c'+T+2)$ per input channel vector representation. This is achieved by considering and learning separate sets of weights, $\{\mathbf{W}_{q}^{(1)},\mathbf{W}_{k}^{(1)}, \mathbf{W}_{v}^{(1)}\}$, for the first attention, and $\{\mathbf{W}_{q}^{(2)},\mathbf{W}_{k}^{(2)}, \mathbf{W}_{v}^{(2)}\}$ for the second attention within a transformer block. The attention is first evaluated for the $i-$th subcarrier over $T$ channel realizations, resulting in an embedding $\mathbf{\widetilde{o}}_i$. Subsequently, to obtain $\mathbf{o}_{i}$, we pass $\mathrm{LayerNorm}(\mathbf{\widetilde{o}}_{i} + \mathbf{\hat{e}}_{i}; \zeta, \iota)$ to the second attention. Attention is evaluated as in equation (\ref{eqn:AttendedEmbedding}). In Fig. \ref{fig:WiTE_time_subcarrier_attention}, we show the performance while varying $T = \{4, 8, 12, 16\}$, and keeping the number of subcarriers constant $N_c' = 32$. While the localization accuracy improves when $T = 8$ (i.e., S32/T8), it starts to degrade for $T>8$. However, we note that as we increase $T$, the available training dataset size is reduced too, hence, resulting in unjust comparison. Nevertheless, we consider S32/T8 model as a tradeoff between the computational complexity and the performance.
	\subsubsection{WiT-E-AF} In contrast to the aforementioned time-frequency separated attention approach, here we view the channel as $\mathbf{H}_r \in \mathbb{C}^{N_c' \times 1 \times N_r}$, and we further \textit{slice} subcarriers into $S$ parts for each antenna element, resulting in an input representation vector $\mathbf{h}_{i} \in \mathbb{R}^{1 \times 3N_c'/S}$. We depict antenna-frequency separated attention in Fig. \ref{fig:WiTE_antenna_subcarrier_attention}, where $\mathbf{\widetilde{o}}_i$ and $\mathbf{o}_{i}$ are obtained similar to that described in WiT-E-TF. The localization performance is shown in Fig. \ref{fig:antenna_frequency_attention} for $N_r = \{4, 8, 16\}$, and $S = 8$. This design choice performs poorly for co-located antennas, whilst gains are clearer in distributed antenna systems as $N_r > 4$.
	\subsubsection{WiT-E-UPA} 
\begin{figure}[!t] 
		\centering 
		\begin{subfigure}{.325\columnwidth}
			\subfloat[WiT-E-TF. \label{fig:time_frequency_attention}]{%
				\includegraphics[width=1.00\linewidth]{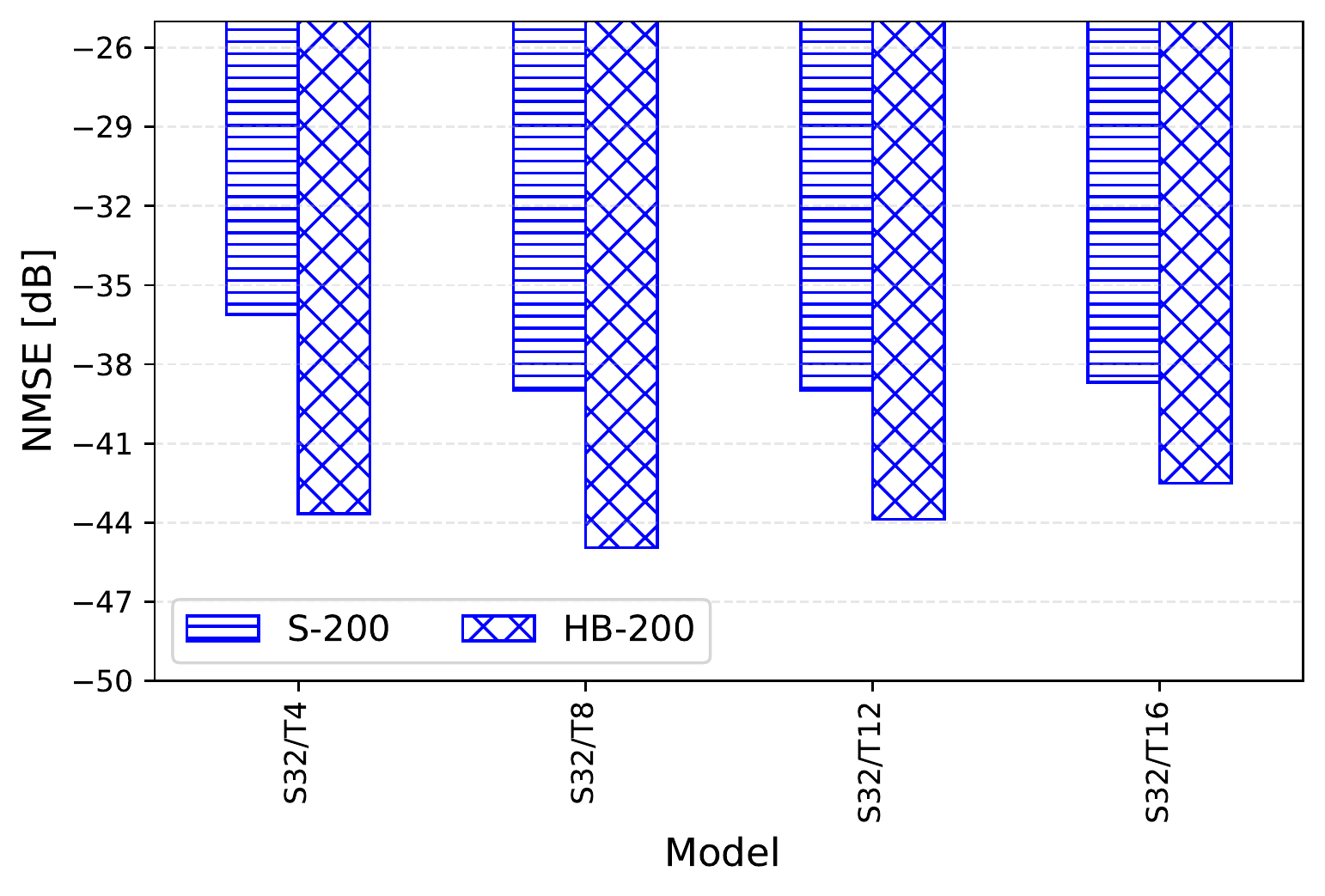}}
		\end{subfigure}
		\hfill
		\begin{subfigure}{.325\columnwidth}
			\subfloat[WiT-E-AF. \label{fig:antenna_frequency_attention}]{%
				\includegraphics[width=1.00\linewidth]{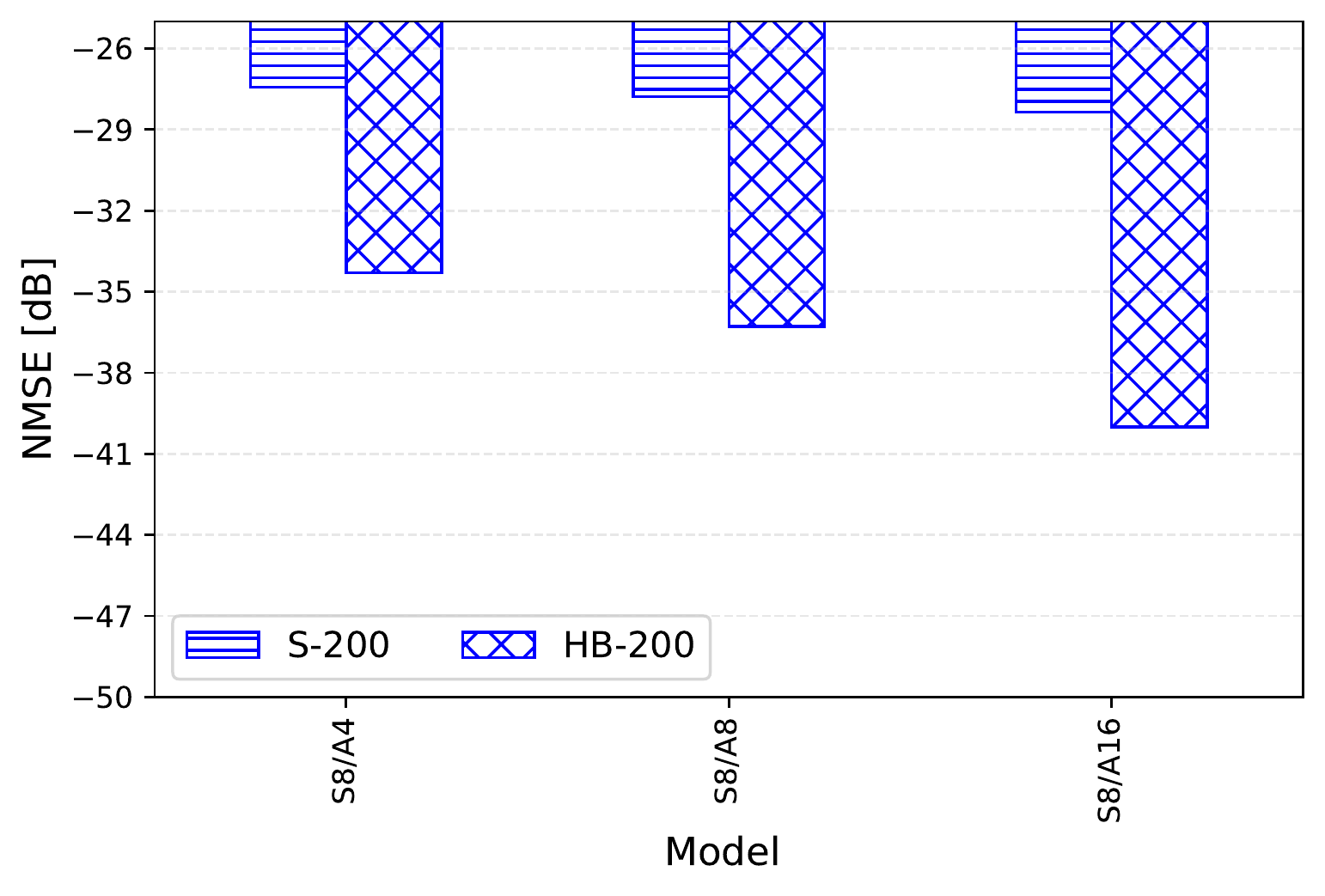}}
		\end{subfigure}
		\hfill
		\begin{subfigure}{.325\columnwidth}
			\subfloat[WiT-E-UPA. \label{fig:subcarrier_antenna_upa_attention}]{%
				\includegraphics[width=1.00\linewidth]{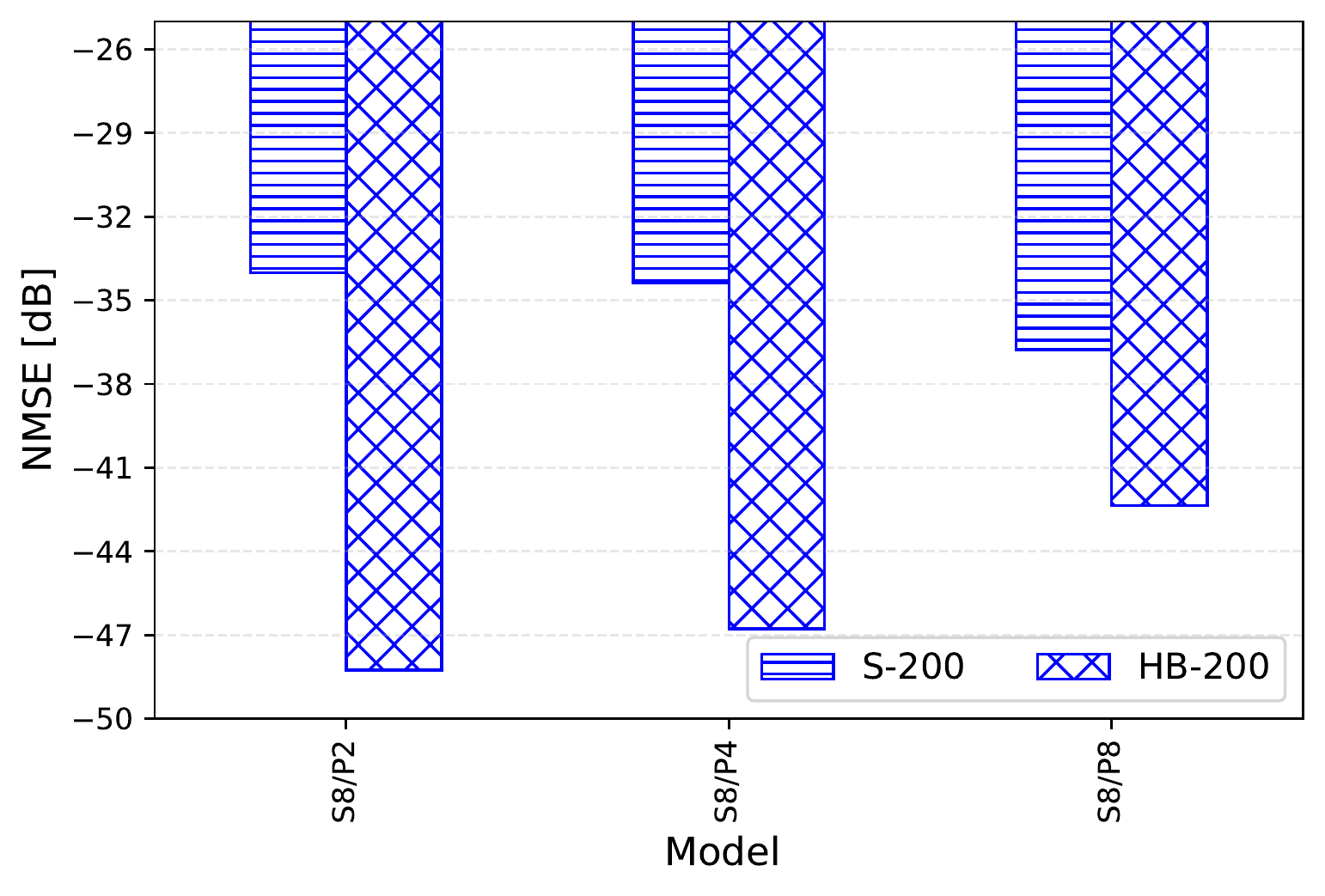}}
		\end{subfigure}
		\caption{Localization performance for three investigated approaches.}
		\label{fig:time_frequency_antenna_separated_atttention}
	\end{figure}

We also attempt to leverage the prior knowledge on the alignment of antenna elements in an array structure or clusters of RRHs. For the co-located antennas, we do so by viewing the channel as $\mathbf{H}_r \in \mathbb{C}^{N_{r_z} \times N_{r_x} \times N_c'}$, where $N_{r_z}$ denotes the number of elements aligned along the height of the array and $N_{r_x}$ along its width. Further, we consider $P = P_1P_2$ sub-arrays, where $P_1$ and $P_2$ are the \textit{slicing} factors across height and width of the array, respectively. The flattened patch for the $i-$th subcarrier results in an input representation vectors $\mathbf{h}_{i} \in \mathbb{R}^{1 \times 3\frac{N_{r_z}}{P_1}\frac{N_{r_x}}{P_2} }$. We depict UPA attention design in Fig. \ref{fig:WiTE_subcarrier_antenna_upa_attention}. Similar to the above elaborated designs, we first evaluate the attention for the channel vector at each sub-array over all subcarriers, then process the \textit{attended} representation through second attention to evaluate over other sub-arrays. The performance is shown in Fig. \ref{fig:subcarrier_antenna_upa_attention} for $P = \{2, 4, 8\}$, and $S = 8$. For DAS, we consider $P$ clusters each having $M/P$ adjoint RRHs. Performance improves in a co-located scenario with $P>2$, but it degrades with increasing number of clusters of RRHs in the case of DAS.
	
	\subsection{Number of Parameters}
	\begin{table}[!b]
		\caption{Input representation and computational complexity.}.
		\centering
		\scalebox{0.95}[0.95]{%
			\begin{tabular}{lccc}
				\Xhline{3\arrayrulewidth}
				\noalign{\vskip\doublerulesep
					\vskip-\arrayrulewidth}
				\Xhline{1\arrayrulewidth} 
				& \multicolumn{3}{c}{S-200 } \\
				\cline { 2 - 4 }
				\shortstack{\textbf{Model}} & $\downarrow$ \shortstack{MAE}  & $\downarrow$ \shortstack{$95-$th} & $\downarrow$ \shortstack{Par. $(\times 10^6)$} \\
				\hline
				\rowcolor{LightCyan}WiT-E-TF S32/T8 & $1.46$ & $4.29$ & $4.78$ \\
				WiT-E-AF S8/A16 & $4.33$ & $13.01$ & $4.66$ \\
				WiT-E-UPA S8/P8 & $1.79$ & $4.45$ & $4.67$ \\
				\rowcolor{LightCyan} WiT $\mathrm{[LID]}$ \cite{9833994} & $2.36$ & $6.54$ & $2.67$ 
				\\
				WiT-L $\mathrm{[LID]}$ & $\mathbf{1.45}$ & $\mathbf{3.78}$ & 	$\mathbf{4.41}$ \\
				\hline 
				\rowcolor{LightCyan} SWiT+Fine-tuning & $1.98$ & $5.66$ & $1.86$ \\
				\\[-0.51cm]				
				\Xhline{3\arrayrulewidth}				
		\end{tabular}}
		\label{table:wit_e_results_table} 
	\end{table}
	In Table \ref{table:wit_e_results_table}, we compare the performance among the proposed approaches in this work for the most performing models. Furthermore, we also show the number of trainable parameters for each model, with only the parameters for the encoder and linear MLP head shown for SWiT+Fine-tuning. Results from Sec. \ref{sub:varyInput} show that incorporating time information improves localization performance for both co-located and distributed antenna systems, but this increases the number of trainable parameters compared to WiT. Thus, to understand the impact of computational complexity, we increase the depth of the WiT (WiT-L), i.e., increase the number of transformer blocks to $H_{\text{blcn}} = 3$, each having $H_{\text{attn}} = 6$ and $D = 128$. By doing so, the number of parameters becomes comparable to WiT-E models, and the accuracy improves (even surpasses them). Finally, in Table \ref{table:wit_e_results_table}, we also compare the self-supervised approach proposed in this work (i.e., SWiT+Fine-tuning). We see that it has lower computational complexity and performs on par with fully-supervised approaches, and even better in some instances. 
	\section{Conclusion}\label{sec:conclusion}
	In this work, we proposed SWiT, a \textit{joint-embedding} self-supervised method for wireless channel representation learning. Our framework is designed to learn invariant and compressed channel representations from both macroscopic and microscopic fading channel characteristics. We showed that the self-supervised approach could outperform other state-of-the-work localization techniques in small-data regimes even under linear evaluation. Additionally, we showed that a pre-trained model on one dataset can be used to assess performance for \textit{spot} estimation in other datasets from different environments or system configurations. Our method also exhibits superior performance when fine-tuning the mapping of channels to absolute location coordinates. Furthermore, we found that the learned representations can achieve highly accurate path-loss predictions using only linear evaluations. Although our approach does not require heavy use of stochastic augmentations, further investigation on the impact of channel transformations is planned in future work.
	
	\bibliographystyle{IEEEtran}
	\bibliography{References}

\end{document}